\documentclass[a4paper,3p,12pt]{elsarticle}

\usepackage{graphicx,multirow}
\usepackage{amssymb}
\usepackage{amsthm}

\usepackage{stmaryrd}

\usepackage{amsmath}
\usepackage{xcolor}
\usepackage{hyperref}
\usepackage{natbib}
\biboptions{sort&compress}
\usepackage{graphicx}
\usepackage{caption}
\usepackage{subcaption}
\usepackage[toc,page]{appendix}

\hypersetup{colorlinks=true}

\newcommand{\figref}[1]{Figure~\ref{#1}}
\newcommand{\tabref}[1]{Table~\ref{#1}}

\newcommand{\bs}[1]{\boldsymbol{#1}}
\newcommand{\norm}[1]{\lVert #1 \rVert}
\newcommand{\pfv}{d} 
\newcommand{\de}[1]{\,{\mathrm d}#1} 

\journal{arXiv.org}

\begin{document}

\begin{frontmatter}

\title{%
    Simulating progressive failure in laminated glass beams with a layer-wise randomized phase-field solver}

\author[dm,dd]{Jaroslav Schmidt}
\ead{jaroslav.schmidt@fsv.cvut.cz}

\author[dg]{Alena Zemanová}
\ead{alena.zemanova@fsv.cvut.cz}

\author[dm]{Jan Zeman$^{*}$\corref{cor}}
\ead{jan.zeman@cvut.cz}

\cortext[cor1]{Corresponding author.}

\address[dm]{Department of Mechanics, Faculty of Civil Engineering, Czech Technical University in Prague, Czech Republic}

\address[dg]{Department of Geotechnics, Faculty of Civil Engineering, Czech Technical University in Prague, Czech Republic}

\address[dd]{Department of Mathematics, Informatics and Cybernetics, University of Chemistry and Technology Prague, Czech Republic}

\begin{abstract}
    Laminated glass achieves improved post-critical response through the composite effect of stiff glass layers and more compliant polymer films, manifested in progressive layer failure by multiple localized cracks. As a result, laminated glass exhibits greater ductility than non-laminated glass, making structures made with it suitable for safety-critical applications while maintaining their aesthetic qualities. However,   such post-critical response is challenging to reproduce using deterministic failure models, which mostly predict failure through a single through-thickness crack localized simultaneously in all layers. This numerical-experimental study   explores the extent to which progressive failure can be predicted by a simple randomized model, where layer-wise tensile strength is modeled by independent, identically distributed Weibull variables. On the numerical side, we employ a computationally efficient, dimensionally-reduced phase field formulation -- with each layer considered to be a Timoshenko beam -- to study progressive failure through combinatorial analysis and detailed Monte Carlo simulations. The reference experimental data were obtained from displacement-controlled four-point bending tests performed on multi-layer laminated glass beams. For certain combinations of the glass layer strengths, results show that the randomized model can reproduce progressive structural failure and the formation of multiple localized cracks in the glass layers. However, the predicted response was less ductile than that observed in experiments, and the model could not reproduce the most frequent glass layer failure sequence. These findings highlight the need to consider strength variability along the length of a beam and to include it in phase-field formulations.
\end{abstract}


\end{frontmatter}

\section{Introduction}\label{S:Intro}

Laminated glass combines stiff glass layers with compliant, transparent polymer interlayers. The interlayers decrease structural stiffness in the pre-peak range but greatly enhance ductility because of the gradual failure of the brittle glass layers. This composite effect enables the deployment of laminated glass in architectural, structural, or automotive engineering for safety-critical applications~(see, e.g.,~\cite{Haldimann:2008:SUG,Bedon2018_psf,Bonenberg2022}) while simultaneously making the mechanics of laminated glass structures intriguing. 

The prevailing design approach to modeling the \emph{pre-fracture behavior} of laminated glass is to replace a heterogeneous structure with a homogeneous one, with the overall thickness adjusted to achieve an equivalent response. This "effective thickness" concept has been successfully implemented, e.g., for deflection~\cite{galuppi2012effective,galuppi2012effectiveP}, buckling~\cite{Lopez-Aenlle201644,D_Ambrosio2020205}, and free vibration~\cite{Aenlle2013,zemanova2018modal} analyses of laminated glass beams and plates. Several analytical studies have also addressed the estimation of failure loads of laminated glass beams~\cite{foraboschi_behavior_2007}, optimal design under mechanical and aesthetic criteria~\cite{foraboschi_optimal_2014}, or the post-breakage stiffness of laminated glass structures under in-plane~\cite{galuppi_homogenized_2016} and out-of-plane~\cite{galuppi_post-breakage_2018} loads. Although analytical approaches provide invaluable insight into the mechanics of laminated glass structures, the detailed \emph{post-fracture response} has mostly been investigated with numerical means, including discrete element~\cite{Zang200773,Baraldi2016278}, cohesive zone~\cite{Chen20161,Vocialta2018448}, and peridynamics~\cite{Wu:2020:OSB,Naumenko2022} formulations~(see also selected reviews~\cite{Wang2017493,Teotia2018412,Kuntsche2019209} for a~detailed comparison). 

In this contribution, we focus on the computational modeling of laminated glass structures using dimensionally-reduced phase-field fracture models. The phase-field approach was first introduced by Bourdin et al.~\cite{bourdin2000numerical} in 2000 as a continuous approximation to the free discontinuity problem arising from Francfort-Marigo's variational reformulation~\cite{Francfort19981319} of Griffith's fracture evolution theory~\cite{griffith1921vi}. Since then, this strategy has gained significant appeal in the computational mechanics community because (i)~no apriori assumptions are made on the crack path, (ii)~crack propagation criteria follow directly from the optimality conditions that replace crack tracking algorithms, and (iii)~the formulation is objective under mesh refinement. Interested readers are referred to~\cite{bourdin2008variational,ambati2015review,Wu2020} for detailed overviews. 

The original phase-field formulations were first developed for two- and three-dimensional bodies, and their adaptation to dimensionally-reduced models applicable to beams, plates, or shells appeared considerably later. Amiri et al.'s~\cite{Amiri2014} pioneering formulation addressed Kirchhoff-Love thin shells with through-thickness cracks and employed maximum-entropy mesh-free approximation to discretize the governing equations. Kiendl et al.~\cite{Kiendl2016} extended this work to account for the tension-compression asymmetry in the thickness direction and based discretization on isogeometric analysis. Further refinements focused on developing formulations for Reissner-Mindlin shells~\cite{Kikis2021}. A more accurate representation of partial fracture along thickness was achieved by using a specific ansatz for the phase field in Euler-Bernoulli beams~\cite{Lai2020} or by employing a separate, dimensionally-reduced representation for kinematic variables and a full representation for the phase field that are coupled at quadrature points~\cite{Ambati2022}. For an illustration of the predictive power of dimensionally-reduced fracture models, see, e.g., a recent study~\cite{Bijaya2023} on fracture propagation in architectured beam lattices.

Several earlier studies utilized a phase-field approach to simulate the post-fracture response of laminated glass structures. Alessi and Freddi obtained the first results for hybrid glass laminates, considering them as elastic-brittle glass layers connected with cohesive interfaces, employing one-dimensional~\cite{ALESSI20179} and two-dimensional~\cite{Alessi2019} models. Freddi and Mingazzi extended this approach to tall laminated glass beams~\cite{Freddi2020}. In our two previous works, we used layer-wise beam and plate models to simulate the response of multi-layered glass samples under quasi-static~\cite{Schmidt:2020:PFF} and low-velocity impact~\cite{Schmidt:2023:PPP} loads.

This article extends our quasi-static study~\cite{Schmidt:2020:PFF} that employed several variants of phase-field formulations combined with 2D plane stress, a layer-wise Reissner-Mindlin plate, and Timoshenko beam structural models to simulate failure in laminated glass structures. The study confirmed that the layer-wise beam theories represent the most suitable options in terms of balancing accuracy and computational requirements. However, all the formulations considered failed to replicate the distributed cracking and the gradual failure sequence observed in the experimental campaign that complemented the numerical investigations. 

\paragraph{Novelty and scope} This work investigates the gradual and distributed cracking in laminated glass structures with randomized phase field models. A well-established Weibull approach~\cite{Weibull1951} is adopted to model the intrinsic variability of the glass strength (see, e.g., a recent review for details) and combined with the Monte Carlo simulations~(see, e.g., related studies~\cite{bonati_redundancy_2019,bonati_probabilistic_2020} on probabilistic modeling based on the Weibull statistics and~\cite{biolzi_estimating_2017,casolo_modelling_2018} on numerical approaches based on the discrete element method or peridynamics). The dimensionally-reduced quasi-static layer-wise beam formulation~\cite{Schmidt:2020:PFF} is used as the deterministic solver to keep the simulation cost manageable. Note that our previous results~\cite{Schmidt:2023:PPP} show that a dynamic phase field formulation can reproduce crack branching in laminated glass under impact loads. However, its computational demands are prohibitive for stochastic simulations and a similar argument holds for the experimental campaign. Besides, because the first steps to the mesh-independent stochastic phase-field fracture approaches have appeared only recently~\cite{hai_relationship_2023,wu_phase-field_2023}, the simplest stochastic model is adopted, which accounts for the variability of the tensile strength in the thickness direction by assuming that layer-wise strengths can be described by independent identically-distributed Weibull random variables.

To this end, the remainder of the paper is organized as follows. Section~\ref{S:PhaseField} introduces the deterministic solver used to predict the response of laminated glass beams, valorizing the main insights from~\cite{Schmidt:2020:PFF}. Section~\ref{sec:testsetup} provides details about the experimental campaign, material parameters for interlayer and glass layers, the setting of the deterministic solver, and its verification using a benchmark problem. Section~\ref{sec:results} collects simulation results, namely a combinatorial analysis to explore the range of responses attainable by the model. It also provides insights into the experimental data inferred from the Monte Carlo simulations. Finally, Section~\ref{sec:conclusions} summarizes the most important findings.

Throughout the text, standard notation is employed, using $a$, $\bs{a}$, and $\bs{A}$ to denote a~scalar quantity, a vector or a second-order symmetric tensor, and a symmetric fourth-order tensor, respectively. The single and double dots indicate the corresponding contractions, e.g., $\bs{a} \cdot \bs{b}$ or $\bs{A} : \bs{b}$, and $\bs{\nabla}$ and $^\prime$ represent the gradient and derivative operators. Other symbols are introduced as needed.

\section{Deterministic phase-field fracture solver}\label{S:PhaseField}

The current section outlines the development of the computational tool used in fracture simulations, adapting the results of our previous study~\cite{Schmidt:2020:PFF} and adopting the following modeling assumptions:

\begin{enumerate}
    \item[(A1)] the sample is subjected to quasi-static loading,
    \item[(A2)] fracture occurs only in the glass layers, 
    \item[(A3)] viscoelastic response of the polymer interlayer is approximated by the quasi-elastic model with time-dependent stiffness, e.g.,~\cite{Galuppi2012,Galuppi2013,ZEMANOVA2017380},
    \item[(A4)] perfect bonding is maintained between glass and polymer layers, e.g.,~\cite[Section~2.3]{chen2017numerical},    
    \item[(A5)] each layer is modeled as a Timoshenko beam~\cite{timoshenko_correction_1921,timoshenko_transverse_1922} with through-thickness cracks. 
\end{enumerate}

\subsection{Model setup}\label{sec:model_setup}

\begin{figure}[!h]
    \centering
    \includegraphics{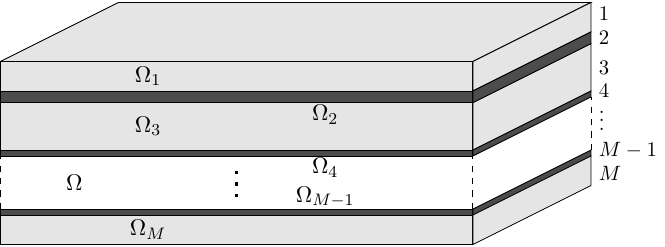}
    \caption{Subdomains in a multi-layer laminated glass beam sample occupying domain $\Omega$. The structure consists of~$M$ layers $\Omega_m$, with the odd indices~($m = 1, 3, \ldots, M$) corresponding to the glass layers and even indices~($m = 2, 4, \ldots, M-1$) to the polymer interlayers.}
    \label{fig:domains}
\end{figure}

We consider the domain of a multi-layer laminated glass sample $\Omega$ to be divided into~$M$ layers $\Omega_m$. Assuming the common setup with an odd number of layers, the odd indices $\{ 1, 3, \ldots, M \}$ correspond to the glass layers, whereas the even indices $\{ 2, 4, \ldots, M - 1\}$ to the polymer interlayers; see \figref{fig:domains} for an illustration. 

Under the assumptions (A1)--(A4), the mechanical response of the laminated structure is encoded in the energy functional 
\begin{align}\label{E:functional}
    \mathcal{E}(t, \bs{u}, \bs{\pfv})
    =
    \sum_{m = 2, 4}^{M-1}
    \mathcal{E}_{m}(t, \bs{u}_m)
    +
    \sum_{m = 1, 3}^{M}
    \mathcal{E}_{m}(t, \bs{u}_m, \pfv_m),
\end{align}
where $t$ denotes the instant time varying over the time interval $[0, T]$, $\bs{u} = (\bs{u}_m)_{m = 1, 2, \ldots, M}$ the displacement fields in all layers, and $\bs{d} = (\bs{d}_m)_{m = 1, 3, \ldots, M}$ the damage phase fields in the glass layers~(the values of $d_m = 0$ and $d_m = 1$ correspond to the intact and fully damage states, respectively). The layer-level energy functionals $\mathcal{E}_m$ have a different structure for glass and polymer layers as specified next. 

For \emph{polymer layers}, the assumptions~(A1)--(A4) imply that their energies follow from
\begin{align}
\mathcal{E}_{m}(t, \bs{u}_m )
=
\Psi^\text{e}_{m}(t, \bs{u}_m )
-
\mathcal{P}_{m}(t, \bs{u}_m),
&&
m = 2, 4, \ldots, M-1,
\label{eq:EF_J}
\end{align}
where $\mathcal{P}_{m}$ stands for the work done by external forces~(if present). 
The elastically stored energy $\Psi^\text{e}_{m}$ is given as
\begin{align}\label{eq:energy_polymer}
    \Psi^\text{e}_{m}(t, \bs{u}_m)
    =
    \int_{\Omega_m}
    \psi^\text{e}_{m}
    \left( 
        t,  \bs{\varepsilon}_m 
    \right)
    \de V,
    && 
    m = 2, 4, \ldots, M-1,
\end{align}
where $\bs{\varepsilon}_m = ( \bs{\nabla} \bs{u}_m + (\bs{\nabla}\bs{u}_m)^\mathsf{T})/2$ designates the small strain tensor and the elastic strain energy density $\psi^\text{e}_{m}$ has the standard quadratic form
\begin{align}
    \psi^\text{e}_{m}(t, \bs{\varepsilon}_m )
    =
    \textstyle{ \frac{1}{2} } \,
    \bs{\varepsilon}_m : 
    \bs{C}_m( G_m(t ), \nu_m ) : 
    \bs{\varepsilon}_m,
    &&
    G_m(t) 
    =
    G_\text{Maxwell}\left( \frac{t}{2}, \theta \right),
\end{align}
with $\bs{\varepsilon} = \bs{\nabla}_\text{s} \bs{u}$ denoting the strain tensor and $\bs{C}_m$ the isotropic stiffness tensor determined from the time-dependent quasi-elastic shear modulus\footnote{%
Note that $G_m$~(the interlayer shear modulus) should not be mistaken with the glass fracture energy $G_m^\text{c}$ introduced later.} $G_m$ and time-independent Poisson ratio $\nu_m$, see~\cite{Duser1999,ZEMANOVA2017380} for further discussion and justification. We assume that the shear modulus can be accurately predicted using the temperature-dependent generalized Maxwell chain model with parameters identified in~\cite{hana2019experimental} are evaluated in the half of the current simulation time according to~\cite{Duser1999}. Because the temperature $\theta$ is considered constant during the whole experiment, we omit the dependence of $G_m$ on this parameter for notational simplicity.

Assumptions~(A1) and~(A2) imply that energies of the \emph{glass layers} attain the form
\begin{align}\label{eq:EF_I}
\mathcal{E}_{m}(t, \bs{u}_m, \pfv_m)
=
\Psi^{\text{e}}_m(\bs{u}_m, \pfv_m)
+ 
\Psi^{\text{d}}_m(\pfv_m)
-
\mathcal{P}_{m}(t, \bs{u}_m),
&& 
m = 1, 3, \ldots, M,
\end{align}
that incorporates the damage variable $d_m$ to account for dissipative processes. In particular, the elastically stored energy in the following form~\cite{miehe2010thermodynamically,pham2011gradient} is considered:
\begin{align}\label{eq:energy_glass}
    \bar{\Psi}^{\text{e}}_{m}(\bs{u}_m, \pfv_m)
    = 
    \int\limits_{\Omega_m} 
    \Bigl(
        (1 - d_m )^2 \, 
        \psi^\textrm{e}_{+, m}( \bs{\varepsilon}_m ) 
        + 
        \psi^\textrm{e}_{-, m}( \bs{\varepsilon}_m )
    \Bigr)
    \de V,
    && 
    m = 1, 3, \ldots, M,
\end{align}
where the split of the energy density into the positive, $\psi^\textrm{e}_{+, m}$, and negative, $\psi^\textrm{e}_{-, m}$, parts is discussed later. Finally, the dissipated energy in the $m$-th layer $\Psi^{\text{d}}_{m}$ is considered in the regularized form proposed in~\cite{pham2011gradient}
 \begin{align}\label{eq:dissipated_energy}
    \Psi^{\text{d}}_{m}(\pfv_m)
    = 
    {\textstyle \frac{3}{8}} \,
    G^{\text{c}}_{m}
    \int\limits_{\Omega_m}
    \Bigl(
    \frac{d_m}{\ell_m} 
    + 
    \ell_m \Vert \bs{\nabla} \pfv_m \Vert^2
    \Bigr)
    \de V,
    && 
    m = 1, 3, \ldots, M,
\end{align}
where $G^{\text{c}}_{m}$ stands for the fracture energy of the $m$-th layer and the regularization parameter $\ell_m$ sets the lengthscale of the damage profile.
 
To track the evolution of the structure over a time interval $[0, T]$, the time interval is discretized into $N$~time instants $0=t_0<t_1<t_2< \cdots <t_N = T$.  . Then, given the initial data $(\bs{u}_0, \bs{d}_0 ) = ((\bs{u}(t_0), \bs{d}(t_0) )$, the displacement and damage fields $(\bs{u}_n, \bs{d}_n ) = ((\bs{u}(t_n), \bs{d}(t_n) )$ at time $t_n$ follow from the recursive energy minimization
\begin{align}\label{eq:incremental_energy_minimization}
    \mathcal{E}( t_n, \bs{u}_n, \bs{d}_n )
    \leq 
    \mathcal{E}( t_n, \bs{u}, \bs{d})
    \text{ for all admissible pairs }
    ( \bs{u}, \bs{d}), 
    &&
    n = 1, 2, \ldots, N.
\end{align}
By the assumption~(A4), the admissible displacement field $\bs{u}$ must remain continuous in~$\Omega$ (and satisfy the kinematic boundary conditions on the boundary $\partial \Omega$ that we did not discuss for the sake of brevity). The admissible damage fields are subjected to the box constraints
\begin{align}\label{eq:irreversibility}
    d_{n-1, m} \leq d_{n, m} \leq 1, 
    &&
    n = 1, 2, \ldots, N; \quad 
    m = 1, 3, \ldots, M,
\end{align}
where the first inequality accounts for the damage irreversibility, i.e., the damage variable cannot decrease in time.

\subsection{Dimensional reduction}\label{sec:dimensional_reduction}
\begin{figure}[h]
    \centering
    \includegraphics[width=.7\textwidth]{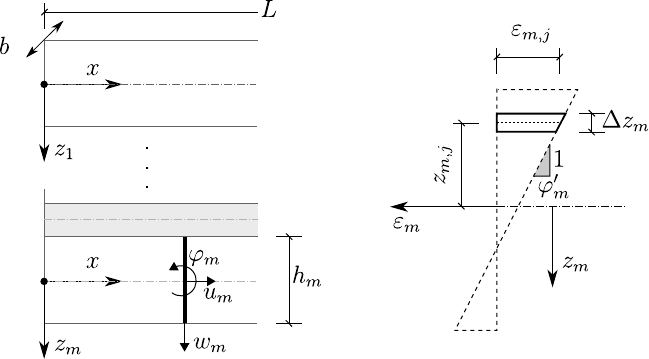}
    \caption{Local coordinate systems, geometrical parameters, and primary unknowns of the layer-wise beam model~(left) and through-thickness distribution of normal strain in the $m$-th layer and the adopted numerical quadrature (right).}
    \label{fig:beam_geom}
\end{figure}

Assumption~(A5) implies the following distribution of normal, $\varepsilon_m$, and shear, $\gamma_m$, strains in the $m$-th layer:
\begin{subequations}
\begin{align}\label{eq:RM_kinematics}
    \varepsilon_m(x,z_m)
    & =
    u^\prime_m(x) 
    + 
    z_m \varphi_m^\prime(x) 
    \,%
    = \varepsilon_{\text{c}, m}(x)
    +
    z_m \kappa_m (x) 
    ,
    \\
    \gamma_m(x) 
    & = 
    \varphi_m(x) 
    + 
    w_m^\prime(x),
    &&
    m = 1, \ldots, M,
\end{align}
\end{subequations}
where $x \in (0, L)$ and $z_m \in (-h_m / 2, h_m / 2)$ denote the position in the layer local coordinate system, $L$ and $h_m$ the beam length and the layer thickness, and $u_m$, $w_m$, and $\varphi_m$ the cross-section horizontal displacement, vertical displacement, and rotation; see also \figref{fig:beam_geom} for illustration. In addition, $\varepsilon_{\text{c}, m}$ designates the centerline strain and $\kappa_m$ the centerline pseudo-curvature. For later reference, we denote the layer (generalized) displacement fields by $\bs{u}_m = (u_m, w_m, \varphi_m)$. 

The rest of the model development follows the same steps as in Section~\ref{sec:model_setup}. For \emph{polymer} interlayers, the elastically stored energy~\eqref{eq:energy_polymer} transforms into
\begin{align}
    \Psi^{\text{e}}_{m}(t, \bs{u}_m)
    = 
    \int_0^L 
    \psi^{\text{e}}_{m}(
        t, \varepsilon_{\text{c}, m}, 
        \kappa_m, 
        \gamma_m )
    \de x,
    &&
    m = 2, 4, \ldots, M-1,
 \end{align}
 with the cross-section elastic energy   
 \begin{align}\label{eq:energy_polymer_red}
    \psi^\text{e}_{m}(t, \varepsilon_{\text{c} ,m}, \kappa_m, \gamma_m )
    = 
    {\textstyle \frac{1}{2} } \,
    G_m( t )  
        \bigl( 
            2(1 + \nu_m ) 
            \left( 
                A_m \varepsilon_{\text{c}, m}^2
                +
                I_m \kappa_m^2
            \right)
            +
            A^*_m \gamma_m^2
        \bigr)
\end{align}
for $m = 2, 4, \ldots, M-1$. Equation~\eqref{eq:energy_polymer_red} additionally involves the layer cross-section characteristics  
\begin{align}
    A_m 
    =
    b h_m,
    &&
    I_m
    = 
    \frac{1}{12} b h^3_m,
    &&
    A^*_m    
    = \frac{5}{6} b h_m,
\end{align}
where $A_m$ stands for the cross-section area, $I_m$ for the second moment of area, and $A^*_m$ for the effective shear area. 

For \emph{glass} layers~($m = 1, 3, \ldots, M$), the elastically stored energy~\eqref{eq:energy_glass} takes the form 
\begin{align}
    \Psi^{\text{e}}_{m}(\bs{u}_m, d_m)
    = 
    \int_0^L
    \Bigl(
    (1 - d_m)^2
    \psi^\text{e}_{+, m}(
        \varepsilon_{\text{c}, m}, 
        \kappa_m, 
        \gamma_m )
    +
    \psi^\text{e}_{-, m}(
        \varepsilon_{\text{c}, m}, 
        \kappa_m, 
        \gamma_m )
    \Bigr)
    \de x,
\end{align}
with the cross-section tensile and compressive energies following from a numerical quadrature in the $z_m$ variable, as first proposed in~\cite{kiendl2016phase}:
\begin{subequations}\label{eq:beam_numerical_integration}
    \begin{align}
    \psi^\text{e}_{+,m}( \varepsilon_{\text{c}, m}, \kappa_m, \gamma_m )    
    & =
    {\textstyle \frac{1}{2}} \,
    E_m b \,
    \sum_{j=1}^J
    \left\langle 
        \varepsilon_{\text{c}, m}
        +
        \kappa_m \, z_{m,j} 
    \right\rangle^2_+ \Delta z_m  
    +
    {\textstyle \frac{1}{2}} \,
    G_m A^*_m \gamma_m^2,
    \label{eq:cross_section_positive}
    \\
    \psi^\text{e}_{-,m}( \varepsilon_{\text{c}, m}, \kappa_m, \gamma_m )    
    & =
    {\textstyle \frac{1}{2}} \,
    E_m b \,
    \sum_{j=1}^J
    \left\langle 
        \varepsilon_{\text{c}, m}
        +
        \kappa_m \, z_{m,j} 
    \right\rangle^2_- \Delta z_m,
    \quad
    m = 1, 3, \ldots, M.
    \label{eq:beam_negative_density}
    \end{align}
\end{subequations}
Here, $E_m$ stands for the Young modulus of the $m$-th layer, $J$ for the number of integration points located at $z_{m,j}$, $\Delta z_m = h_m / J$ their integration weight (see \figref{fig:beam_geom}), and $\langle \bullet \rangle_+$ and $\langle \bullet \rangle_-$ for the positive and negative parts of $\bullet$. The regularized dissipated energy, on the other hand, follows directly from the original expression~\eqref{eq:dissipated_energy}:
\begin{align}\label{eq:dissipated_energy_1D}
\Psi^\text{d}_{m}( \pfv_m )
= 
{\textstyle \frac{3}{8}} \,
G^\text{c}_m A_m
\int_0^L
\Bigl(
    \frac{d_m}{\ell_m}
    +
    \ell_m
    \left( d_m^\prime \right)^2
\Bigr)
\de x,
&&
m = 1, 3, \ldots, M
\end{align}
because the damage variable is assumed to be constant over a cross-section; recall assumption~(A5). 

Having reformulated all the components of the model introduced in Section~\ref{sec:model_setup}, the evolution of the system follows from the same incremental energy minimization problem~\eqref{eq:incremental_energy_minimization} under damage irreversibility enforced by~\eqref{eq:irreversibility}. The interlayer displacement continuity implies that 
\begin{subequations}\label{eq:continuity_conditions}
    \begin{align}
    w_m & = w_1, && m = 2, 3, \ldots, M, 
    \label{eq:rest_w} 
    \\
    u_{m-1} 
    + 
    {\textstyle \frac{1}{2}}
    \varphi_{m-1} h_{m-1}  
    & 
    = 
    u_{m} 
    - 
    {\textstyle \frac{1}{2}}
    \varphi_{m} h_{m},
    &&
    m = 2, 3, \ldots, M.
    \label{eq:rest2}
    \end{align}
\end{subequations}
Therefore, the centerline horizontal displacements $u_m$ and cross-section rotations $\varphi_m$ in the polymer layers can be expressed using the respective quantities in the glass layers:
\begin{subequations}\label{eq:rest}
    \begin{align}
    u_m 
    & = 
    {\textstyle \frac{1}{4}}\,
    \left(
        h_{m-1}\varphi_{m-1} 
        - 
        h_{m+1}\varphi_{m+1}
    \right) 
    + 
    {\textstyle \frac{1}{2}}\,
    \left(
        u_{m-1} 
        + 
        u_{m+1}
    \right), 
    &&
    m = 2, 4, \ldots, M-1,
    \label{eq:rest_u}
    \\
    \varphi_m 
    & = 
    \frac{1}{h_m}
    \left( 
        u_{m+1} 
        - 
        u_{m-1}
        -
        {\textstyle \frac{1}{2}}\,
        \varphi_{m-1} h_{m-1} 
        -
        {\textstyle \frac{1}{2}}\,
        \varphi_{m+1}h_{m+1}  
    \right), 
    &&
    m = 2, 4, \ldots, M-1.
    \label{eq:rest_phi}
    \end{align}
\end{subequations}
For instance, the kinematics of a $5$-layer laminated glass beams ($3$ glass layers and $2$ interlayers) is specified by $7$ independent fields: $w_1, u_1, \varphi_1, u_3, \varphi_3, u_5$, and $\varphi_5$.    

\subsection{Weak form of governing equations}

The equations governing the evolution of displacement and damage fields in the layer-wise beam model follow from the optimality conditions of the variational problem~\eqref{eq:incremental_energy_minimization}. In particular, the optimality conditions with respect to the kinematic quantities yield: 
\begin{subequations}\label{E:min_u}
\begin{align}
\bs{\nabla}_{\bs{u}_m}
\Psi^\text{e}_{m}(\bs{u}_{n,m}, d_{n,m})
\cdot
\delta\bs{u}_{m}
-
\bs{\nabla}_{\bs{u}_m}
\mathcal{P}_m (t_n, \bs{u}_{n,m} ) 
& = 0,
&& 
m = 1, 3, \ldots, M,
\label{E:min_u_glass}
\\
\bs{\nabla}_{\bs{u}_m}
\Psi^\text{e}_{m}(t_n, \bs{u}_{n,m})
\cdot
\delta\bs{u}_{m}
-
\bs{\nabla}_{\bs{u}_m}
\mathcal{P}_m (t_n, \bs{u}_{n,m} ) 
& = 0,
&& 
m = 2, 4, \ldots, M - 1,
\label{E:min_u_interlayer}
\end{align}
\end{subequations}
where both the solution $\bs{u}_{n,m}$ at time $t_n$ and an admissible variation $\delta\bs{u}_{m}$ satisfy the continuity conditions~\eqref{eq:continuity_conditions}. For reader's convenience, the full form of the derivatives $\bs{\nabla}_{\bs{u}_m}
\Psi^\text{e}_{m}$ is collected in~\ref{app:energy_derivatives}.

Because of the irreversibility constraints on the damage variables~\eqref{eq:irreversibility}, the respective optimality conditions attain the form of a variational inequality  
\begin{align} \label{E:min_d}
    2 
    \int_0^L
    \left( d_{n,m} - 1 \right)
    Y_{n,m}    
    \delta{\pfv}_m
    \de x 
    +
    {\textstyle \frac{3}{8}} \,
    G^\text{c}_{m} A_m
    \int_0^L
    \Bigl(
    \frac{\delta d_m}{\ell_m}
    +
    2
    \ell_m d^\prime_{n, m} \delta d^\prime_m
    \Bigr)
    \de x
    \geq 
    0,
\end{align}
where $m = 1, 3, \ldots, M$ and the admissible variations $\delta d_m$ satisfy $d_{n-1, m} \leq d_{n, m} + \delta d_m \leq 1$, cf.~\cite{Jirasek:2015:LSR}. The quantity $Y_{n,m}$ abbreviates the damage driving force obtained as 
\begin{align}
    Y_{n, m}( x )
    =
    \psi^\text{e}_{+, m}\left(
        \varepsilon_{\text{c}, m} (x), 
        \kappa_m (x), 
        \gamma_m (x) 
        \right),
    &&
    x \in (0, L);
\end{align}
recall~\eqref{eq:cross_section_positive} for the definition of the tensile cross-section energy $\psi^\text{e}_{+, m}$. 

Our previous study~\cite{Schmidt:2020:PFF} revealed that the most efficient and robust phase field solver is obtained when replacing the variationally consistent driving force in Eq.~\eqref{E:min_d} with  
\begin{eqnarray}\label{eq:mod_en}
    Y_{n,m}(x)
    \approx
    \textstyle{\frac{1}{2}} \,
    E_m A_m 
    \max 
    \bigl(
        \langle\varepsilon_m(x, h_m/2)\rangle_+^2,
        \langle\varepsilon_m(x, -h_m/2)\rangle_+^2 
    \bigr),
    && x \in (0, L), 
\end{eqnarray}
where the normal strain in the $m$-th layer $\varepsilon_m$ follows from the displacement field $\bs{u}_{n,m}$ using the parameterization~\eqref{eq:RM_kinematics}  . This choice effectively introduces a Rankine-type failure condition, as first proposed in~\cite{miehe2015phase}; see also~\cite{Bilgen2019} and Section~\ref{sec:phase-field_params} for further discussion. Notice that the modified version of the driving force $Y_{n, m}$ enters only the damage evolution equations~\eqref{E:min_d}, whereas the original form of the tensile energy $\psi^\text{e}_{+, m}$ from Eq.~\eqref{eq:cross_section_positive} is used in the equilibrium equations~\eqref{E:min_u}.

\subsection{Implementation}\label{S:implementation}

The set of two governing equations~\eqref{E:min_u} and~\eqref{E:min_d} is resolved with the Finite Element Method (FEM), employing a Python interface to the FEniCS library~\cite{logg2012automated} that allows an automatic treatment of the corresponding weak forms. This section briefly outlines the implementation strategy; the full details are available in the accompanying GitLab repository~\cite{git_article}. 

In particular, we employ the \emph{staggered} approach to resolve the two optimality conditions sequentially. Following~\cite{bourdin2000numerical},   the kinematic quantities are first updated from~\eqref{E:min_u} with the damage fields fixed to the values from the previous iteration. Because of the non-linearities induced by the positive-negative split~\eqref{E:min_u_glass_full}, the resulting system of non-linear equations is solved with the Newton-Raphson method with a relative termination tolerance $10^{-12}$. The second step involves the update of the damage fields using~\eqref{E:min_u} while freezing the kinematic quantities to the ones from the first step. For this step, we employ the FEniCS-SNES solver based on a semi-smooth Newton method for variational inequalities. The iterations terminate when the staggered error in the $i$-iteration
\begin{align}\label{E:error}
    \xi^{(i)}
    =
    \max
    \Bigl\{
        \frac{\norm{\bs{w}^{(i)} - \bs{w}^{(i-1)}}_2}
        {\norm{\bs{w}^{(i)}}_2}
        , 
        \frac{\norm{\bs{d}^{(i)} - \bs{d}^{(i-1)}}_2}
        {\norm{\bs{d}^{(i)}}_2}
        \Bigr\}
\end{align}
drops below the tolerance set as $\mathrm{tol}=10^{-6}$. Here, column matrices $\bs{w}^{(i)}$ and $\bs{d}^{(i)}$ collect the vertical deflections and damage variables in all finite element nodes in the $i$-th iteration. 

\begin{figure}[h]
    \centering
    \includegraphics{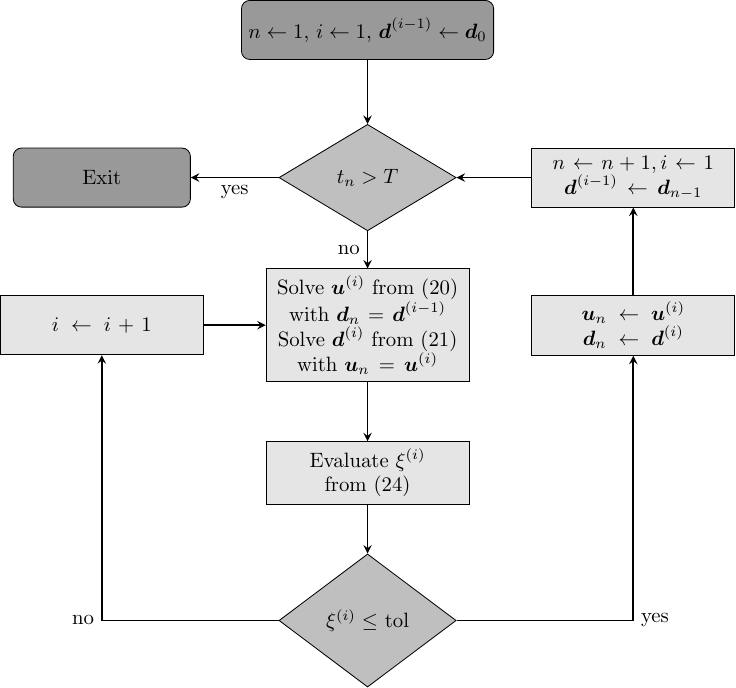}
    \caption{Flowchart of the staggered static phase-field algorithm.}
    \label{fig:flowchart}
\end{figure}

An interested reader is referred to \figref{fig:flowchart} for a flowchart of the algorithm and to, e.g.,~\cite{ambati2015review,Wu2020,Dujc2021} for more details on the staggered approach.

\section{Test and solver setup}\label{sec:testsetup}

In this section, the structural-scale testing data (in Section~\ref{sec:experiments}), the determination of material parameters (Section~\ref{sec:material_parameters}), the setting of the finite element model (Section~\ref{sec:phase-field_params}), and its verification using a single-layer benchmark~(Section~\ref{sec:benchmark}) are collected to provide the basis for the validation studies presented later. Whenever relevant, the observed or simulated failure sequences are compared with analytical approaches based on event tree analyses~\cite{bonati_redundancy_2019}.

\subsection{Experimental campaign}\label{sec:experiments}

\begin{figure}[!h]
    \centering
    \includegraphics{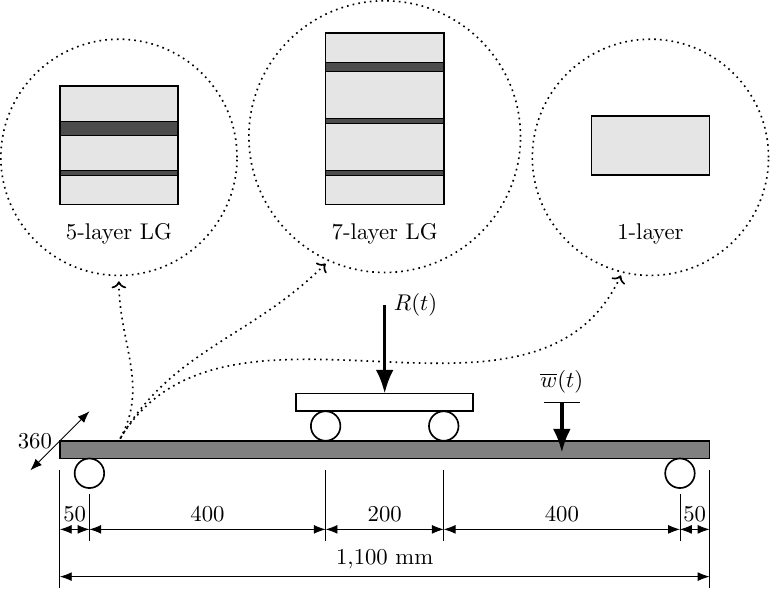}
    \caption{Scheme of the four-point bending test on multi-layer laminated glass beams. $\overline{w}(t)$ denotes the time evolution of the prescribed displacement and $R(t)$ the corresponding reaction.}
    \label{fig:experiment}
\end{figure}

The experiments were carried out by our colleagues from the Experimental Centre and Department of Steel and Timber Structures at the Faculty of Civil Engineering, Czech Technical University in Prague. 
Similarly to our previous study~\cite{Schmidt:2020:PFF}, the four-point bending setup was considered with the arrangement indicated in \figref{fig:experiment} for one 5-layer and two 7-layer laminate configurations specified in \tabref{tab:samples}.

\begin{table}[ht]
    \centering
    \small
    \begin{tabular}{lccc}
         \hline
         & 5LG & 7LG-1 & 7LG-2 \\
         \hline
         number of layers & 5 & 7 & 7 \\
         layer thicknesses [mm] & 5/2.28/6/0.76/5 & 5/1.52/8/0.76/8/0.76/5 & 6/1.52/6/0.76/6/1.52/6\\
         overall thickness [mm] & 20.04 & 29.04 & 27.8 \\
         \hline
    \end{tabular}
    \caption{Composition of laminated glass samples and nominal dimensions.}
    \label{tab:samples}
\end{table}

Laminated glass samples were placed into loading device and central deflections were measured by two displacement sensors. Following the standardized procedure specified in the EN1288-3~\cite{EN1288-3}, the glass specimens were protected by rubber pads at contact with the loading and supporting steel cylinders to prevent stress concentration in these areas. All tests were displacement-controlled 
with a constant loading rate $\dot{\overline{w}}$ of 1 mm/min until the specimen failure. The prescribed displacements $\overline{w}(t)$, for which we determine the time evolution of the reaction force $R(t)$, increased monotonously in time without any loading/unloading cycles after the progressive failure of individual layers. Additional details on the tests are available in~\cite{Hana2018:FPB,Hana2020:FPB,konrad_laminated_2022}.

Three specimens of each beam sample specified in \tabref{tab:samples} were tested. \tabref{tab:exp_data} additionally provides the ambient temperatures during the test and the failure sequence leading to the structural collapse. The data acquired for the third specimen of sample 7LG-1 were defective and hence are excluded from further discussion.

\begin{table}[ht]
    \centering
    \small
    \begin{tabular}{lccc}
         \hline
         sample & specimen & temperature [$^\circ$C] & failure sequence \\
         \hline
          5LG & 1 & 23.0 & 5 $\rightarrow$ 3 $\rightarrow$ 1\\ 
              & 2 & 23.4 & 5 $\rightarrow$ 3 $\rightarrow$ 1\\ 
              & 3 & 23.1 & 5 $\rightarrow$ 3 $\rightarrow$ 1\\ 
          7LG-1 & 1 & 23.3 & 5 + 7 $\rightarrow$ 3 $\rightarrow$ 1\\ 
                & 2 & 23.3 & 5 + 7 $\rightarrow$ 3 $\rightarrow$ 1\\ 
                & 3 & \multicolumn{2}{c}{defective measurement} \\
          7LG-2 & 1 & 28.4 & 7 $\rightarrow$ 5 $\rightarrow$ 1 + 3 \\  
                & 2 & 29.4 & 7 $\rightarrow$ 1 + 3 + 5  \\ 
                & 3 & 27.7 &  7 $\rightarrow$ 5 $\rightarrow$ 3 $\rightarrow$ 1 \\ 
         \hline
    \end{tabular}
    \caption{Overview of glass samples and specimens, ambient temperature during the tests, and the failure sequence. Glass layers are indexed according to \figref{fig:domains}.}
    \label{tab:exp_data}
\end{table}

\begin{figure}[h]
    \centering
    \includegraphics{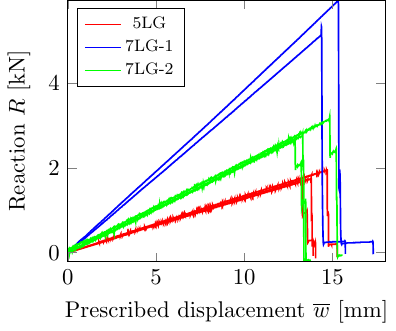}
    \caption{Force-displacement diagrams for all laminated glass beam samples and specimens. Solid lines are used to distinguish different samples.}
    \label{fig:exps}
\end{figure}

\begin{figure}[p]
    \begin{subfigure}{\textwidth}
        \centering
        \includegraphics[width=.75\textwidth]{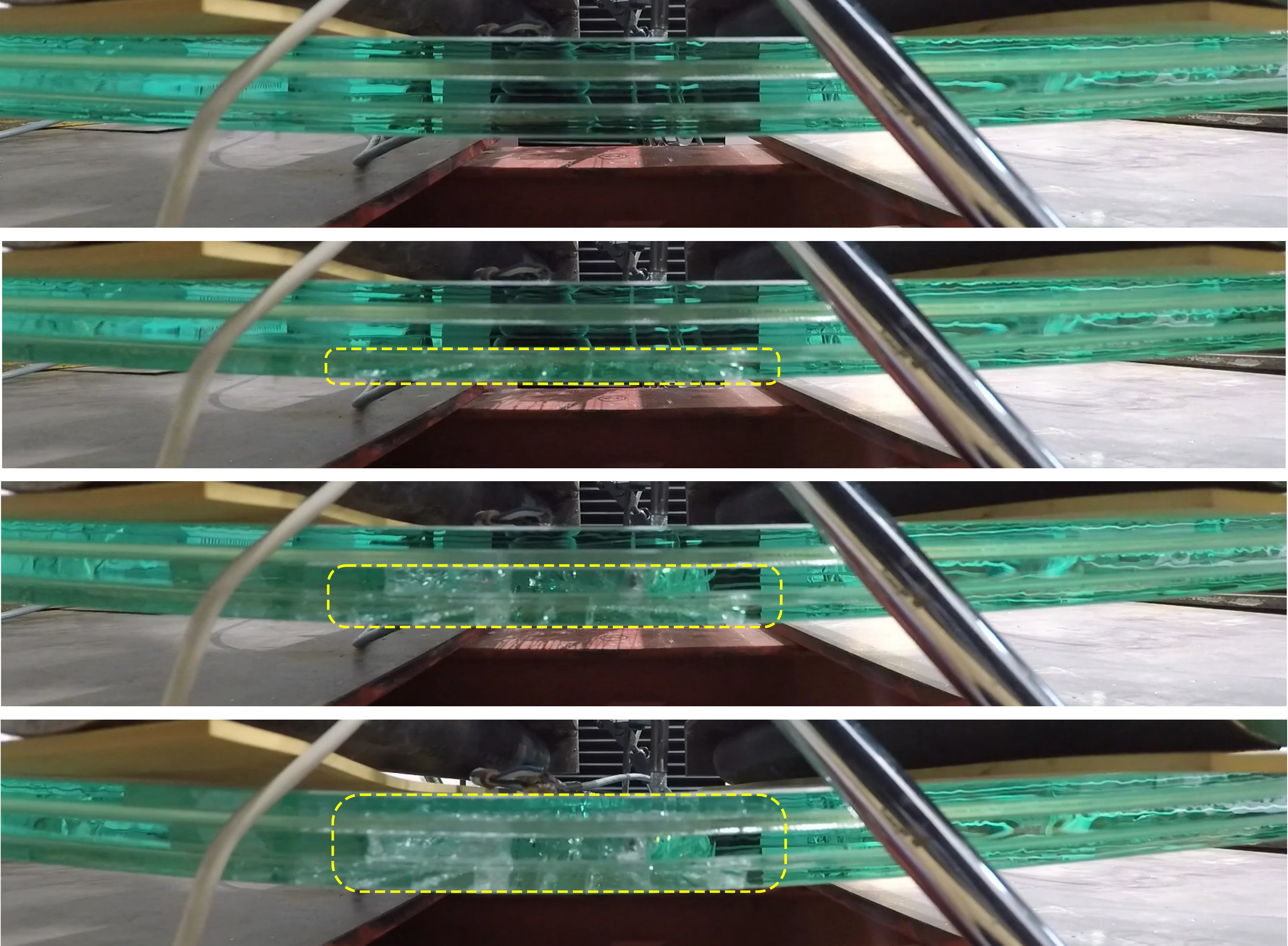}
        \caption{}
        \label{fig:sub_fracture_5LG}
    \end{subfigure}
    \begin{subfigure}{\textwidth}
        \centering
        \includegraphics[width=.75\textwidth]{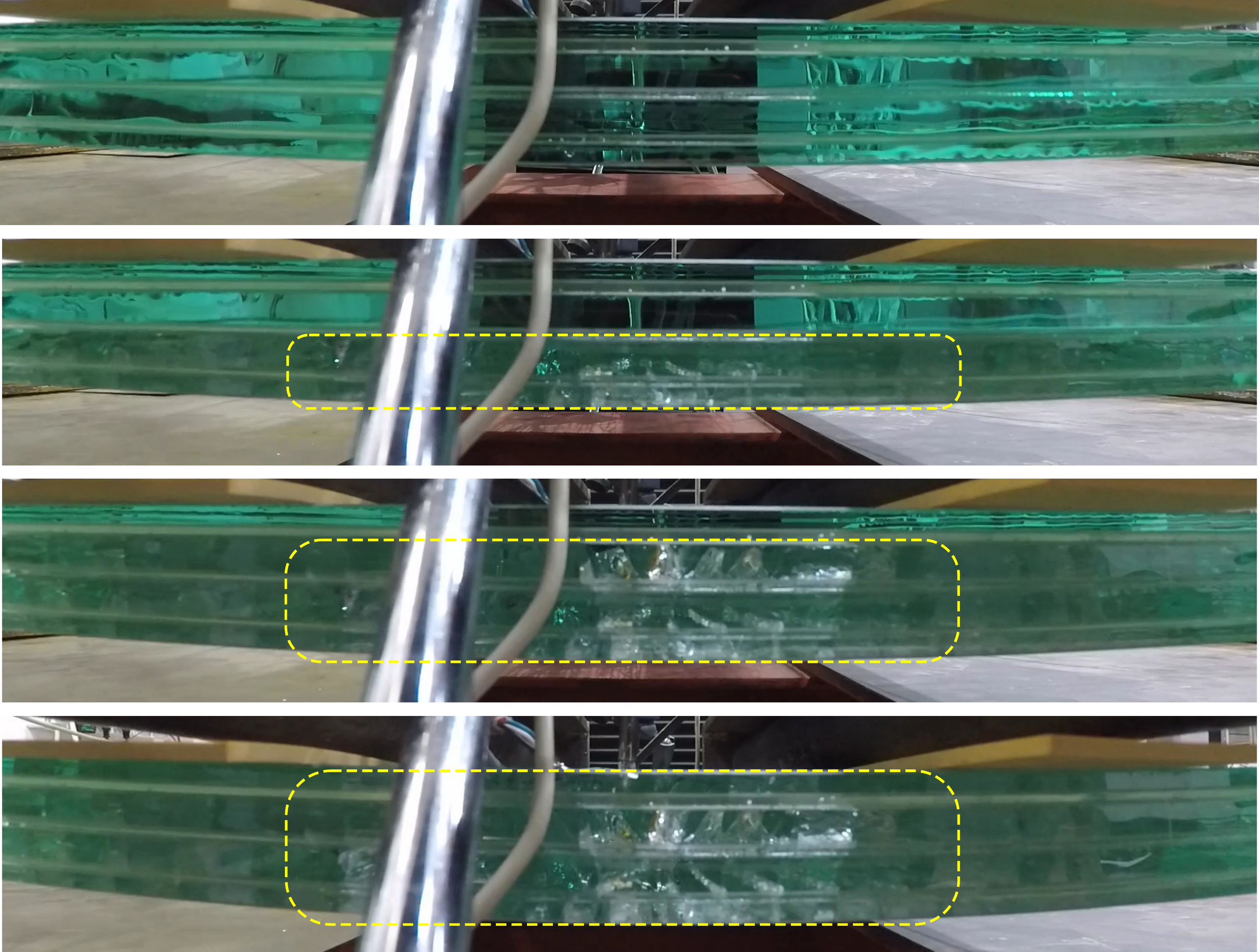}
        \caption{}
        \label{fig:sub_fracture_7LG1}
    \end{subfigure}
    \caption{Progressive failure of glass layers in (a)~specimen 1 of sample 5LG (failure sequence: $5 \rightarrow 3 \rightarrow 1$) and (b)~specimen 1 of sample 7LG-1 (failure sequence: $5 + 7 \rightarrow 3 \rightarrow 1$). The layers are indexed according to \figref{fig:domains}.}
    \label{fig:sub_fracture}
\end{figure}

\begin{figure}[h]
    \centering
    \includegraphics{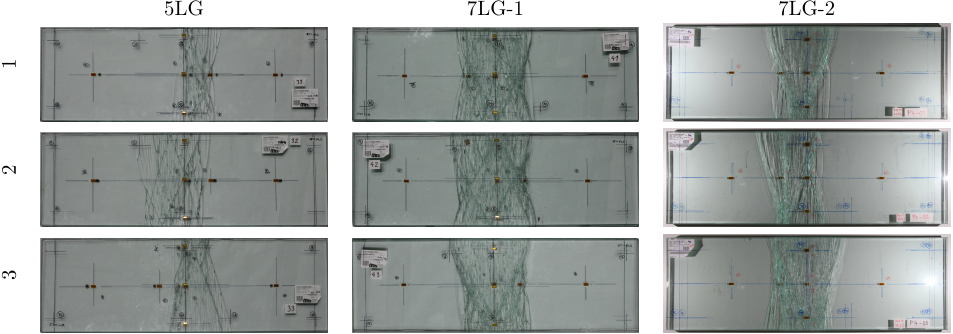}
    \caption{Top view of final fracture patterns on tested laminated glass beams. Row labels correspond to samples, and column labels to specimens.}
    \label{fig:final_pattern}
\end{figure}

The load-displacement diagrams for all samples appear in \figref{fig:exps}. Note that because of the indentation of steel cylinders into the rubber pads, the measured data included an initial non-linear branch that is omitted in the results presented in the next section. The gradual fracture of glass layers manifests in the load-displacement curves by the jumps in the reaction force at a constant displacement of the loading head; see also \figref{fig:sub_fracture} for an illustration of the crack development along the thickness. This information is complemented by the fracture patterns at the sample failure, collected in \figref{fig:final_pattern}, that document the development of distributed cracks in individual layers.\footnote{%
Clearly, the multiple curved crack paths visible in \figref{fig:final_pattern} cannot be reproduced by the adopted layer-wise beam models that assume straight cracks along the beam width. However, as shown in~\cite[Sections 4.3 and 4.4]{Schmidt:2020:PFF}, this modeling assumption leads to nearly identical load-displacement curves and failure characteristics compared to plate formulations, even when explicitly introducing an initial flaw to trigger the development of inclined cracks.}

Notice that in all cases, the failure process initiated in the bottom layer(s) and progressed upward. From the point view of the fault tree considerations~\cite{bonati_redundancy_2019}, this suggests that the degree of shear coupling induced by interlayers was sufficient to ensure that the structural response corresponded more to the monolithic limit (in which the bottom layers carry the largest tensile stresses and, therefore, exhibit higher failure probabilities) than to the layered limit with independent layer responses and thus same layer failure probabilities.

\subsection{Material parameters}\label{sec:material_parameters}

In all the considered laminated glass structures, PolyVinyl Butyral~(PVB) is used as an interlayer material and equivalent elastic response is determined from the temperature-depended generalized Maxwell model with parameters obtained in~\cite{hana2019experimental} by an inverse analysis procedure and provided in the repository~\cite{git_article}. Uniform but different temperatures were assumed for each simulation, as specified in \tabref{tab:samples} on page~\pageref{tab:samples}. 

All glass layers are made of annealed float glass, with the standard Young modulus $E=70$~GPa and Poisson ratio $\nu=0.22$,~\cite{haldimann2008structural}. To characterize the variability of the glass tensile strength $f_\text{t}$, we combined the data collected on monolithic glass samples in~\cite{veer2009} (24 samples \textcolor{red}{under displacement control tests at a loading rate of 1.0 mm/min}) with our previous independent experiments~\cite{Zemanova2018} on three-layer laminated samples (10 samples);
see the repository~\cite{git_article} for the full data set. Each sample in the set had the same geometry, loading type and rate, and type of edge treatment. Four-point bending tests were conducted using a MTS testing machine. Note that for this type of test, the strength exhibited by the glass sample is mainly governed by the damage caused by the edge finishing process instead of flaws randomly distributed on the surface~\cite{Castori2019}. For a detailed discussion of the effect of edge treatment on edge strength, see, e.g., the following studies~\cite{braun2020cut, Bukieda2020}.

\begin{figure}[h]
    \centering
    \includegraphics{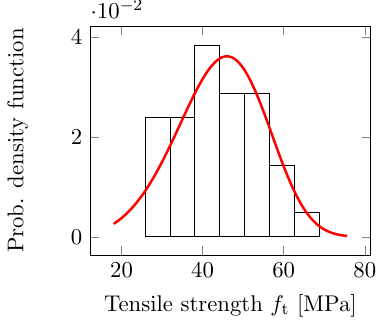}
    \includegraphics{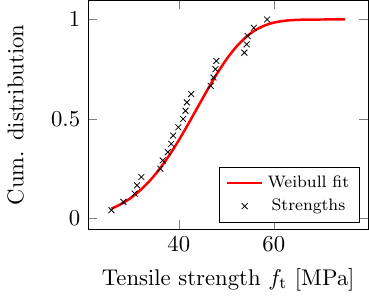}
    \caption{Histogram of tensile strengths complemented with the Weibull fit (left) and corresponding probability plot (right).}
\label{fig:Weibull}
\end{figure}

These data were fitted with the Weibull distribution~\cite{Weibull1951} with the shape parameter $k=4.64$ and the scale parameter $\lambda=48.47$:
\begin{align}\label{eq:weibull}
    f_\text{t}\sim\text{Weibull}(k=4.64, \lambda=48.47)~\text{MPa};    
\end{align}
see \figref{fig:Weibull} for the quality of the fit.
The corresponding $5\%$ or $95\%$ quantiles follow from the strength cumulative distribution function $F_{f_\text{t}}$ as:
\begin{align}\label{eq:quantiles}
    f_{\mathrm{t, lo}} 
    = 
    F_{f_\text{t}}^{-1}(0.05) \doteq 25.6~\text{MPa}, 
    &&
    f_{\mathrm{t, hi}} 
    = F_{f_\text{t}}^{-1}(0.95) \doteq 61.4~ \text{MPa},
\end{align}
and these values are considered to be the extreme values used in the deterministic simulations later in Section~\ref{sec:results}. Note that all the input data were determined from glass samples of a similar length as in our setup to eliminate the stochastic size effect, e.g.,~\cite[Chapter~6]{Bazant:2021} for a general exposition and~\cite{bonati_redundancy_2019,bonati_probabilistic_2020} for dedicated discussion for laminated glass structures. The tensile strength values, therefore, should be understood as effective layer-wise quantities.

\subsection{Phase-field model and spatial discretization}\label{sec:phase-field_params}

As shown in~\cite{wu2018length}, the regularized form of the dissipated energy~\eqref{eq:dissipated_energy_1D} introduces an implicit coupling between the fracture energy~$G^\text{c}$, the tensile strength~$f_\mathrm{f}$, and the characteristic length $\ell$ in the form:
\begin{align}\label{eq:transform_gc_ft}
    G^\text{c}
    =
    \frac{8}{3}\frac{f_\text{t}^2 \ell}{E}.
\end{align}
In our simulation, the value of the layer-wise fracture energy is thus determined according to this relation, assuming the   distribution of the tensile strength~\eqref{eq:weibull} and the characteristic length $\ell = 2h_\mathrm{e}$, where $h_\mathrm{e}$ stands for the uniform element length. 

The value of $h_\mathrm{e} = 0.5$~mm yields a reasonable compromise between the adequate resolution for sharp cracks and the efficiency needed when generating many stochastic realizations. Because of the latter reason,   only a symmetric half of the beam was considered in simulations and employed $J = 40$ integration points for the evolution of energies $\psi^\text{e}_{+,m}$ in Eq.~\eqref{eq:beam_numerical_integration} in the thickness direction. For the discretization along the beam length, we used the mesh of $1,100$ elements per layer and employed linear basis functions for each unknown field $u_m$, $v_m$, $\varphi_m$, and $d_m$ (with one-point reduced integration to eliminate the shear locking). Overall, this setting resulted in typical times in the order of minutes for simulating a single experiment described in Section~\ref{sec:experiments} on a conventional laptop with the Intel(R) Core(TM) i7-8550U processor, with larger simulation times needed for seven-layer samples exhibiting distributed cracking.

\subsection{Verification benchmark}\label{sec:benchmark}

\begin{figure}[!h]
    \centering
    \begin{tabular}{cc}
        \includegraphics[width=.315\textwidth]{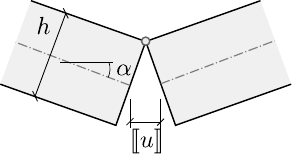} &
        \includegraphics[width=.315\textwidth]{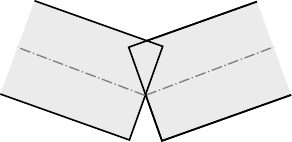} \\
        \small (a) & \small (b)
    \end{tabular}
    \caption{Post-fracture kinematics of single-layer glass beam (a)~without and (b)~with cross-section overlapping.}
    \label{fig:bench}
\end{figure}

We consider a monolithic single-layer problem to verify the implementation of the solver, in particular, the cross-section non-penetration condition after fracture. The pre-fracture behavior of the single-layer beam setup from \figref{fig:experiment} is governed by the elementary theory of thin beams, e.g., \cite[Chapter~4]{Gross:2018:EM2}, which predicts the following relation between the prescribed displacement $\overline{w}$ and the maximum normal stress in the beam $\sigma_{\max}$: 
\begin{align}
    \overline{w}
    =
    \frac{l^2}{3 E h} 
    \left( 
        3 \left( \frac{a}{l} \right) 
        -
        4 \left( \frac{a}{l} \right)^2
    \right)
    \sigma_{\max},
\end{align}
where $E = 70$~GPa is the Young modulus of glass~\cite{Haldimann:2008:SUG}, $h = 20$~mm is the beam height, $l = 1,000$~mm denotes the beam span, and $a = 400$~mm stands for the location of the loading cylinders relative to supports; recall \figref{fig:experiment}. Setting the maximum stress to the mean value of the tensile strength $\sigma_{\max} = f_{\mathrm{t}, \mathrm{mean}} = 45$~MPa according to Eq.~\eqref{eq:weibull} yields the prescribed displacement at failure $\overline{w}_\mathrm{f} = 6$~mm. After the fracture, the structure behaves as the mechanism consisting of two rigid bodies connected by an eccentric hinge, see \figref{fig:bench}a, leading to the jump of the horizontal displacement at the mid-span    
\begin{align}
    \llbracket u \rrbracket = h \sin \alpha, 
    &&
    \alpha = \arctan \frac{\overline{w}_\mathrm{f}}{a};
\end{align}
for the current data, we obtain $\alpha \doteq 1.500 \cdot 10^{-2}$~rad and $\llbracket u \rrbracket \doteq 0.300$~mm.

The same problem was simulated using the finite element model, with the Young modulus of the central finite element decreased by $0.1 \%$ to ensure that a single crack localized at the mid-span. We obtained the values of $\overline{w}_\mathrm{f}^\mathrm{FE} \doteq 6.006$~mm and $\llbracket u \rrbracket^\mathrm{FE} \doteq 0.3003$~mm. Among others, these results confirm the cross-section overlapping has been eliminated by the positive-negative split of the stored energy~\eqref{eq:beam_negative_density}, without which $\llbracket u \rrbracket^\mathrm{FE} = 0$~mm; see \figref{fig:bench}b.

\section{Results}\label{sec:results}

The present section collects the results of numerical simulations of the progressive failure of laminated glass samples introduced in Section~\ref{sec:experiments}. We depart from validating our implementation of a single-layer beam benchmark with a closed-form solution in Section~\ref{sec:benchmark}. 
Section~\ref{sec:mean_value} briefly discusses the results with mean strength data, while Section~\ref{sec:combinatorial} assesses the range of low- and high-strength layers and the corresponding failure behavior. Finally, in Section~\ref{sec:stochastic}, we collect the results of Monte-Carlo simulations and compare them with the available experimental data.

\subsection{Mean value analysis}\label{sec:mean_value}

\begin{figure}[h]
    \centering
    \includegraphics{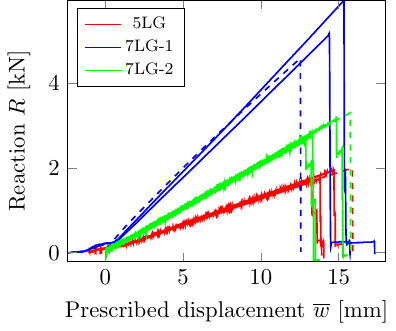}
    \caption{Force-displacement diagrams for all laminated glass beam samples and specimens. Solid lines show experimental results, and dashed lines show simulated results with average tensile strength in glass layers.}
    \label{fig:exps_with_simulation}
\end{figure}

We begin our discussion by briefly comparing experimentally determined load-displacement curves with simulation data for deterministic glass layer strengths with mean values $f_\mathrm{t, mean} \doteq 45$~MPa~(recall Section~\ref{sec:material_parameters}). The results shown in~\figref{fig:exps_with_simulation} confirms that the initial linear response is well-captured by the numerical model, while the maximum displacement $\overline{w}$ and reaction force $R$ are overestimated for samples 5LG and 7LG-2 and underestimated for sample 7LG-1. More importantly, all models predict perfect elastic-brittle failure through simultaneous fracture of all glass layers by a single crack. Such behavior is consistent with our previous study~\cite{Schmidt:2020:PFF}.

\subsection{Combinatorial analysis}\label{sec:combinatorial}

To investigate whether the difference in failure modes among individual specimens and distributed cracking observed in Section~\ref{sec:experiments} can be explained by variable strength distribution, this section collects the mechanical response of all $2^3$ and $2^4$ combinations of the strengths $f_\mathrm{t,lo}$ and $f_\mathrm{t,hi}$ from Eq.~\eqref{eq:quantiles} for the samples 5LG in Section~\ref{sec:5-layer}, and 7LG-1 in Section~\ref{sec:7-layer}, respectively~(the results for the sample 7LG-2 are similar thus omitted for the sake of brevity).

\subsubsection{5-layer beam}\label{sec:5-layer}

\begin{figure}
    \centering
    \begin{subfigure}{.495\textwidth}
        \centering
        \includegraphics[width=\textwidth]{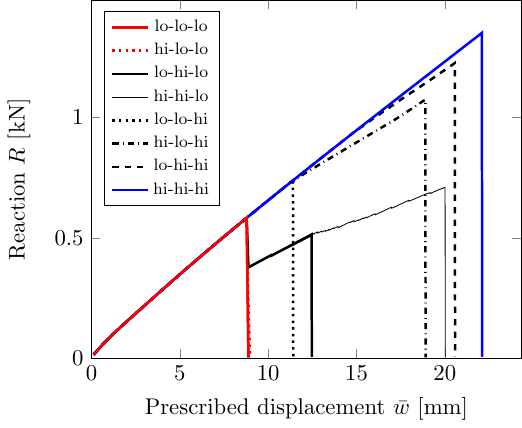}
        \caption{}
        \label{fig:5L_det_force}
    \end{subfigure}
    \bigskip
    \begin{subfigure}{.495\textwidth}
        \centering
        \includegraphics[width=\textwidth]{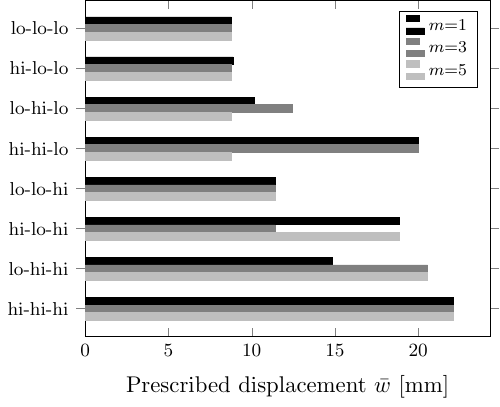}
        \caption{}
        \label{fig:5L_det_integrity}
    \end{subfigure}

    \caption{Evolution of (a)~reaction $R$ and (b)~layer integrity as a function of the prescribed displacement $\overline{w}$ for all combinations of minimum and maximum glass layer strengths in the 5-layer beam sample 5LG. The layers are ordered in the top-to-bottom direction and enumerated by an index $m$ according to \figref{fig:domains}.}
    \label{fig:5L_det}
\end{figure}

All simulation results for the sample 5LG are collected in \figref{fig:5L_det} and are visualized using the force-displacement diagrams (\figref{fig:5L_det_force}) and bar charts showing the integrity of individual glass layers until failure (\figref{fig:5L_det_integrity}). The particular combinations are distinguished by the distribution of the low and high strengths in the top-to-bottom direction, according to \figref{fig:domains}. For example, 'lo-hi-hi' combination indicates a structure with the top glass layer of strength $f_\mathrm{t,lo}$ and the middle and bottom layers of strength $f_\mathrm{t,hi}$.

\begin{figure}[p]
    \centering
    \begin{tabular}{ccccc}
    \multicolumn{5}{c}{\footnotesize lo-lo-lo} \\
    \includegraphics[height=0.25\textwidth]{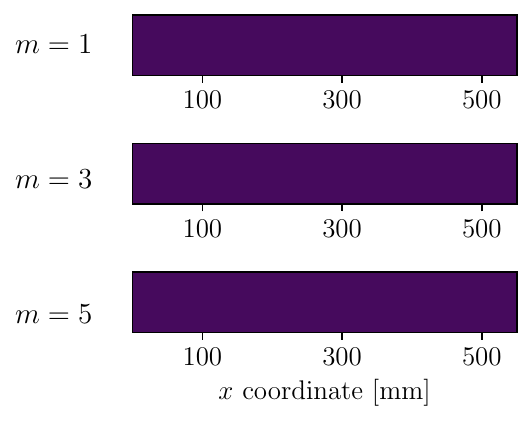} & 
    {\raisebox{20mm}{$\rightarrow$}} & 
    \includegraphics[height=0.25\textwidth]{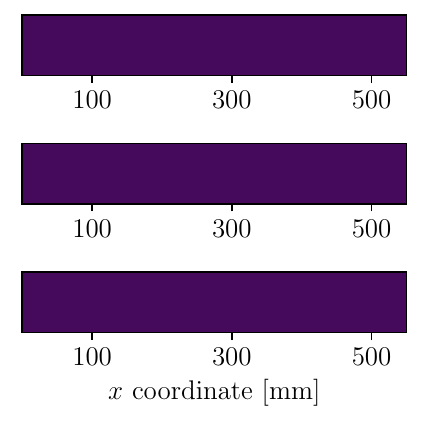} & 
    {\raisebox{20mm}{$\rightarrow$}} & 
    \includegraphics[height=0.25\textwidth]{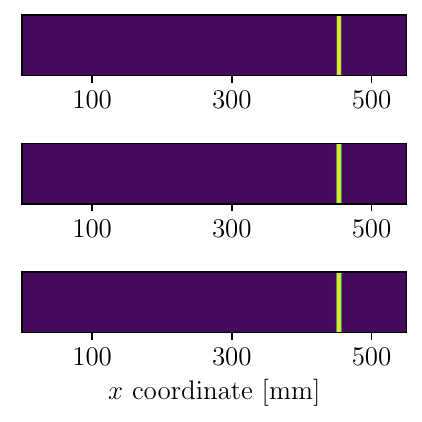} \\
    \hline 
    \multicolumn{5}{c}{\footnotesize lo-hi-lo} \\
    \includegraphics[height=0.25\textwidth]{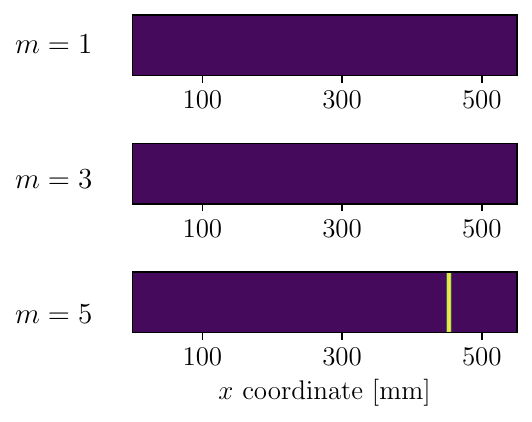} & 
    {\raisebox{20mm}{$\rightarrow$}} & 
    \includegraphics[height=0.25\textwidth]{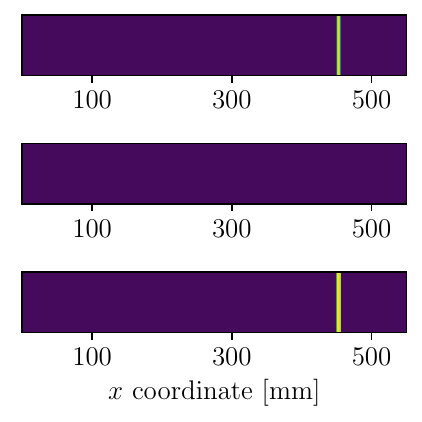} & 
    {\raisebox{20mm}{$\rightarrow$}} & 
    \includegraphics[height=0.25\textwidth]{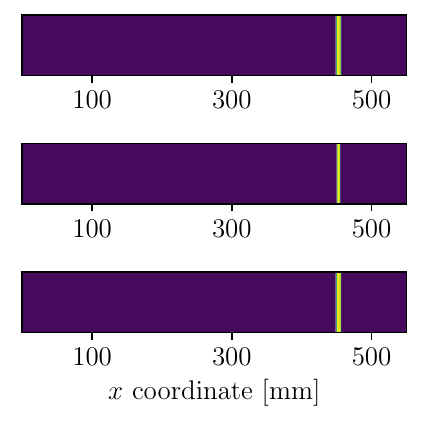} \\
    \hline 
    \multicolumn{5}{c}{\footnotesize hi-hi-lo} \\
    \includegraphics[height=0.25\textwidth]{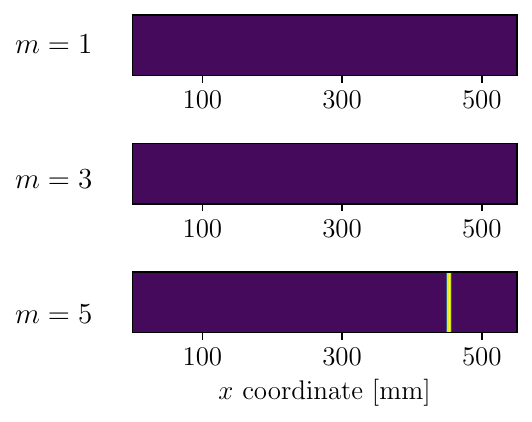} & 
    {\raisebox{20mm}{$\rightarrow$}} & 
    \includegraphics[height=0.25\textwidth]{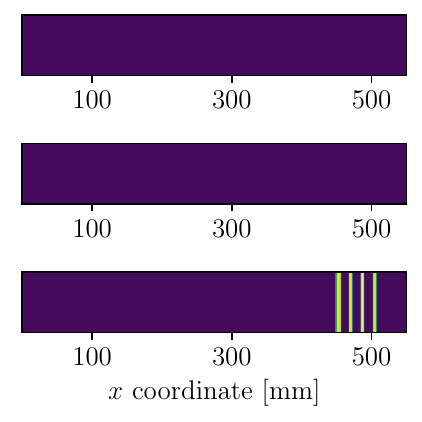} & 
    {\raisebox{20mm}{$\rightarrow$}} & 
    \includegraphics[height=0.25\textwidth]{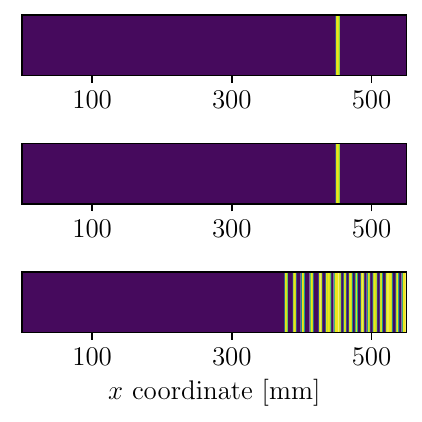}
	\end{tabular}
    \caption{Crack development for three representative combinations of minimum and maximum glass layer strengths: lo-lo-lo, lo-hi-lo, and lo-hi-hi in 5-layer beam sample 5LG. The violet color corresponds to an undamaged cross-section ($d=0$) and yellow to a fully fractured cross-section ($d=1$). The layers are ordered in the top-to-bottom direction and enumerated by an index $m$ according to \figref{fig:domains}.}
    \label{fig:5L_damage_evolution_det}
\end{figure}

As expected, the simulated responses remain confined between the uniform combinations of low (red curve in \figref{fig:5L_det_force}) and high (blue curve) strengths for which the crack propagates through all three layers simultaneously;  the same behavior is observed for the lo-lo-hi combination. These results correspond to our previous findings in~\cite{Schmidt:2020:PFF}. All the remaining combinations exhibit progressive collapse and gradual crack development in individual layers, although the difference from the sudden failure is only minor for the hi-lo-lo combination. We particularly highlight the lo-hi-hi, hi-hi-lo, and lo-hi-lo combinations that may represent the situation in which one or both outer glass layers are unintentionally scratched during transportation and manipulation. Nevertheless, their failure behavior differs. For the lo-hi-lo combination, scratched from both sides, the cracks first propagate through the bottom layer ($m=5$). The top layer ($m=1$) fractures under an additional load increase, followed by structural failure after the rupture of the middle layer ($m=3$). A different behavior is observed for configurations with a single low-strength layer adjacent to the top or bottom surface. Here, the first crack appears in the low-strength layer. Still, upon an increasing load, an array of equally spaced cracks starts to develop until the remaining two layers fail by the simultaneous formation of a single crack per layer. The lo-hi-lo combination exhibits similar behavior, except the fragmentation remains confined to the middle layer. The fracture evolution for the selected configurations is additionally shown in \figref{fig:5L_damage_evolution_det}.

\subsubsection{7-layer beam}\label{sec:7-layer}

\begin{figure}
    \centering
    \begin{subfigure}{.495\textwidth}
        \centering
        \includegraphics[width=\textwidth]{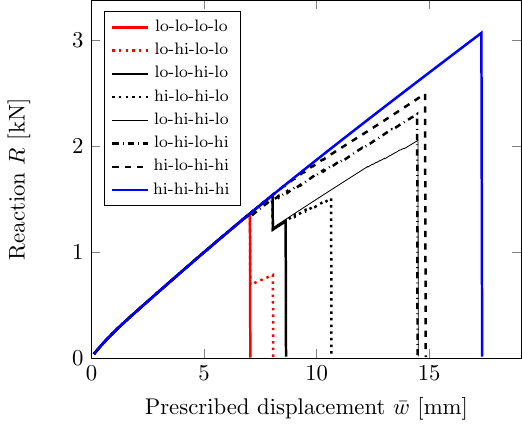}
        \caption{}
        \label{fig:7L_det_force}
    \end{subfigure}
    \begin{subfigure}{.495\textwidth}
        \centering
        \includegraphics[width=\textwidth]{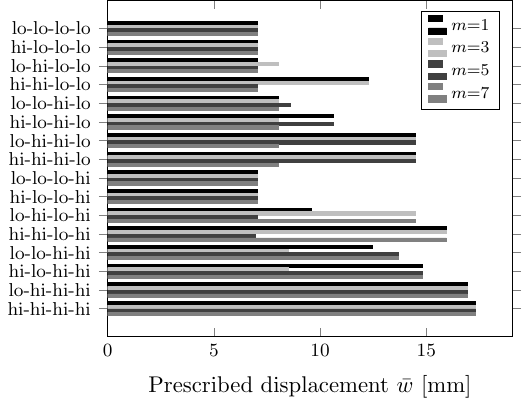}
        \caption{}
        \label{fig:7L_det_integrity}
    \end{subfigure}
    \caption{Evolution of (a)~reaction $R$ and (b)~layer integrity as a function of the prescribed displacement $\overline{w}$ for the (a)~selected and~(b) all combinations of minimum and maximum glass layer strengths in the 7-layer beam sample 7LG-1. The layers are ordered in the top-to-bottom direction and enumerated by an index $m$ according to \figref{fig:domains}.}
    \label{fig:7L_det}
\end{figure}

An analogous analysis for the 7-layer beam 7LG-1 reveals the same mechanisms of the failure processes, as shown for the selected layer combinations in \figref{fig:7L_det_force} and all layer combinations in \figref{fig:7L_det_integrity}. 

\begin{figure}
    \centering
    \begin{tabular}{ccccc}
    \multicolumn{5}{c}{\small lo-hi-lo-hi} \\
    \includegraphics[height=0.26\textwidth]{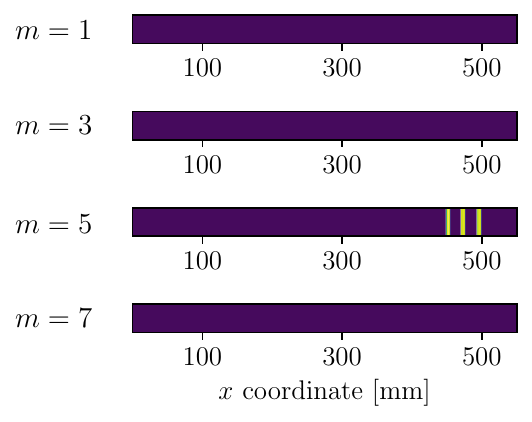} & 
    {\raisebox{20mm}{$\rightarrow$}} & 
    \includegraphics[height=0.26\textwidth]{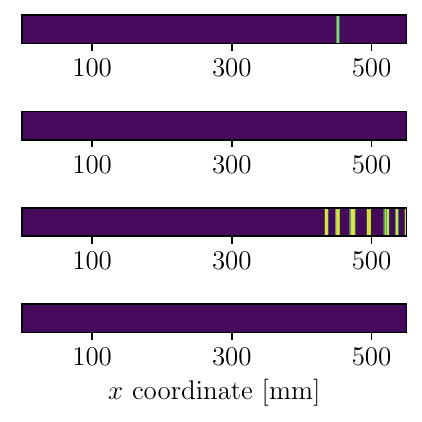} & 
    {\raisebox{20mm}{$\rightarrow$}} & 
    \includegraphics[height=0.26\textwidth]{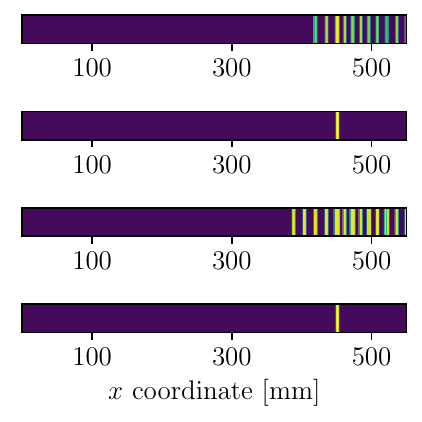} 
    \end{tabular}
    \caption{Crack development for the lo-hi-lo-hi combination of glass layer strengths in the 5-layer laminated glass beam~(sample 7LG-1). The violet color corresponds to an undamaged cross-section ($d=0$) and yellow to a fully fractured cross-section ($d=1$). The layers are ordered in the top-to-bottom direction and enumerated by an index $m$ according to \figref{fig:domains}.}
    \label{fig:7L_damage_evolution_det}
\end{figure}

In particular, the uniform strength distribution through the beam thickness leads to a brittle failure by a single crack aligned in all glass layers. The same brittle response occurs for configurations lo-hi-hi-hi, hi-lo-lo-hi, lo-lo-lo-hi, and hi-lo-lo-lo with a uniform strength distribution in the adjacent glass layers; cf. \figref{fig:7L_det_integrity}. The alternating distributions of minimum and maximum layer strengths lead to a more progressive collapse and even the formation of equally-spaced cracks in specific glass layers, as illustrated in \figref{fig:7L_damage_evolution_det}.

\subsection{Stochastic analysis}\label{sec:stochastic}

The results of the combinatorial analysis confirmed the numerical model's potential to predict the progressive cracking phenomena; this section, therefore, deals with the extension to randomized simulations. Note that the results for all samples are presented simultaneously because, according to the combinatorial analysis, 5- and 7-layer samples revealed similar failure mechanisms. Recall that the outcome of the experimental campaign in Section~\ref{sec:experiments} involved the measured force-displacement diagrams~(shown in \figref{fig:exps}) with failure sequences~(\tabref{tab:exp_data} and Figures~\ref{fig:sub_fracture} and~\ref{fig:final_pattern}). These data are compared here with the results of 1,000 Monte-Carlo simulations, in which the layer strengths were sampled from the Weibull distribution displayed in \figref{fig:Weibull}.  

\begin{figure}[h!]
    \centering
    \includegraphics{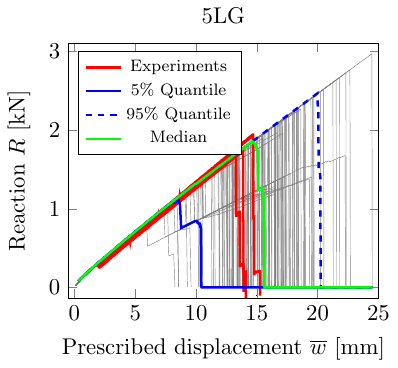}
    \includegraphics{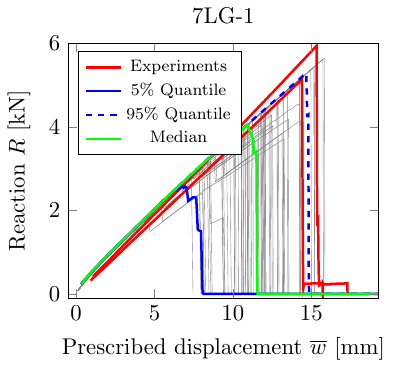}
    \includegraphics{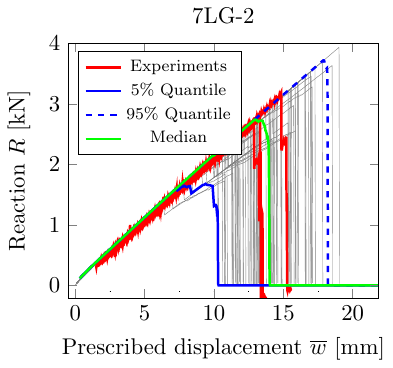}
    \caption{Results and statistics reaction-displacement curves obtained from $1,000$ Monte-Carlo~(MC) simulations of laminated glass beam samples 5LG, 7LG-1, and 7LG-2. The thick red curves correspond to experimental data. Thin gray curves denote individual realizations, blue curves the mean values, and the green curves the upper and lower $5~\%$ quantiles determined from the MC simulations.} 
    \label{fig:stoch_reaction_force}    
\end{figure}

The simulated force-displacement diagrams are shown in \figref{fig:stoch_reaction_force} with thin gray lines, accompanied by the experimental results with thick red lines. The range of predicted responses is substantial and aligns well with the results from the combinatorial analysis conducted in Section~\ref{sec:combinatorial} as seen by comparing \figref{fig:stoch_reaction_force} with Figures~\ref{fig:5L_det_force} and~\ref{fig:7L_det_force}. To analyze the simulated data,   the reaction values $R$ were sampled for a set of prescribed displacements $\overline{w}$ at 1/30 mm intervals. For each $\overline{w}$ value, we constructed an empirical distribution function from the complete set of 1,000 Monte-Carlo simulations, from which   the median values (shown in \figref{fig:stoch_reaction_force} as green curves) and the $5\%$ and $95\%$ quantiles (blue solid and dashed curves) were extracted. Note that the $5\%$~quantile response exhibits gradual cracking, and the $95\%$ quantile transitions to the brittle response. For samples 5LG and 7LG-2, the experimental response falls between the two quantiles, and the median response approximates the experimental data well. For sample 7LG-1, the median response underestimates the experimental response, which is close to the $95\%$ quantile curve. However, the smoothing nature of the median response (especially near the peak)  and the consideration of extreme value statistics over all realizations disregard the information about the failure sequences, and more refined interpretations of the Monte-Carlo data are needed.   

\begin{figure}[h]
    \centering
    \includegraphics{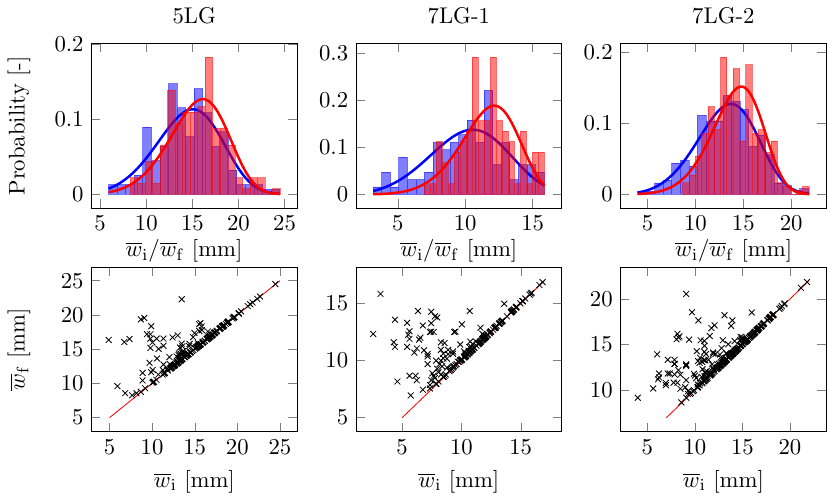}
    \caption{Top: Histograms of displacements inducing the first crack $\overline{w}_\mathrm{i}$~(blue) and the final crack $\overline{w}_\mathrm{f}$~(red) displacements and their Weibull fits for the 5-layer sample 5LG and 7-layer samples 7LG-1 and 7LG-2. Bottom: Plots of the final $\overline{w}_\mathrm{f}$ against the first $\overline{w}_\mathrm{i;}$ crack displacements.}
    \label{fig:stoch_hist}
\end{figure}

For this purpose, let us analyze displacements $\overline{w}_\mathrm{i}$ leading to the formation of the initial cracked layer~(corresponding to the occurrence of a fully-cracked cross-section with $d=1$ in at least a single layer) and $\overline{w}_\mathrm{f}$ leading to the complete structural failure (the occurrence of a fully-cracked cross-section at all layers). The resulting data for all three samples are visualized in \figref{fig:stoch_hist} using two means. The first one is based on histograms of the initial crack displacements $\overline{w}_\mathrm{i}$~(in blue) and the complete failure $\overline{w}_\mathrm{f}$~(in red), along with their fits with the Weibull distributions~\eqref{eq:weibull}. The fitted distributions display visible differences between the two quantities as the ratios between the most likely values (modes) $\overline{w}^\mathrm{mod}_\mathrm{i} / \overline{w}^\mathrm{mod}_\mathrm{f}$ reach 0.93~(for sample 5LG), 0.93~(7LG-1), and 0.91~(7LG-2). However, a different conclusion follows from the bottom graphs in \figref{fig:stoch_hist}, in which we plotted the failure displacement against the displacements at the initial failure for all realizations. These graphs show that for most realizations $\overline{w}_\mathrm{i} = \overline{w}_\mathrm{f}$, indicating the brittle failure by crack(s) running through all glass layers.

\begin{figure}[h]
    \centering
    \includegraphics[width=0.45\textwidth]{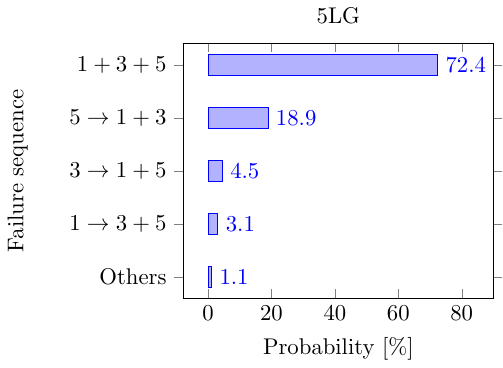}
    \includegraphics[width=0.45\textwidth]{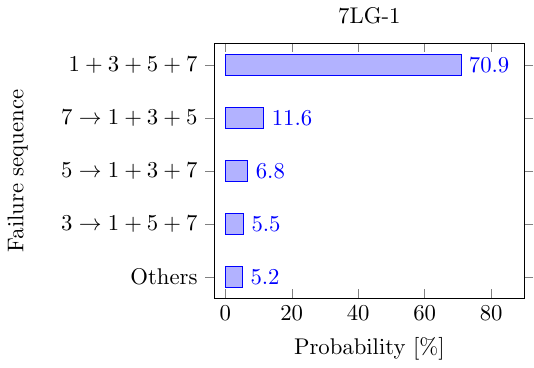}
    \includegraphics[width=0.45\textwidth]{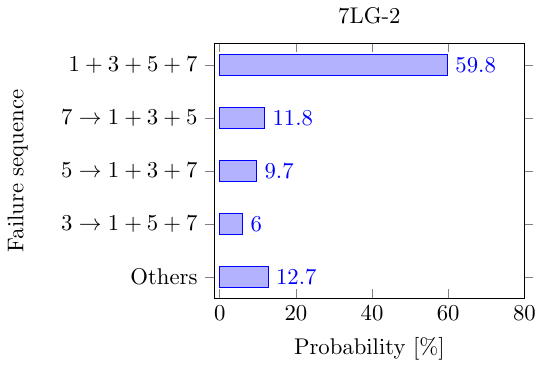}
    \caption{Probability of occurrence of different failure sequences for the 5-layer sample 5LG and 7-layer samples 7LG-1 and 7LG-2. The layers are ordered in the top-to-bottom direction.}
    \label{fig:failure_sequence_probability}
\end{figure}

This claim is further substantiated by \figref{fig:failure_sequence_probability}, showing the probabilities of occurrence for the three most frequent failure sequences. The most common failure scenario for all samples involves simultaneous failure for all layers, with a higher likelihood for the 5-layer sample than for the 7-layer configurations. The second most likely failure sequence proceeds from the collapse of the bottom layer, followed by the simultaneous failure of all remaining layers. This scenario is the closest to the ones observed in experiments for all specimens, in which the failure initiated in the bottom layer   gradually propagated, layer-by-layer, to the top surface under small displacement increments (exceptionally through the simultaneous failure of two layers); recall Section~\ref{sec:experiments} for additional details. Hence, the response predicted by the proposed model is less ductile due to neglecting variability in material properties along the lengths of individual layers; see~\cite{biolzi_estimating_2017,casolo_modelling_2018} for related results in this direction.

Similarly to the experimental results, the numerical simulations confirm that the mechanical response of the laminated beam is much closer to the monolithic than to the laminated limit failure-wise~\cite{bonati_redundancy_2019}. In particular, the scenarios corresponding to the failure initiation in the two bottommost layers account for more than $80\%$ of the simulated cases ($95.8\%$ for 5LG, $89.3\%$ for 7LG-1, and $81.3\%$ for 7LG-2), cf. \figref{fig:failure_sequence_probability}.

\begin{figure}[h]
    \centering
    \includegraphics[width=0.4\textwidth]{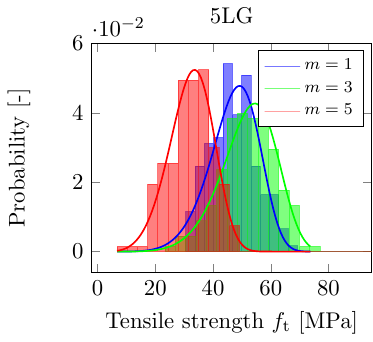}
    \includegraphics[width=0.4\textwidth]{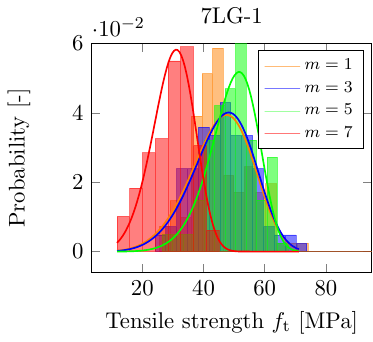}
    \includegraphics[width=0.4\textwidth]{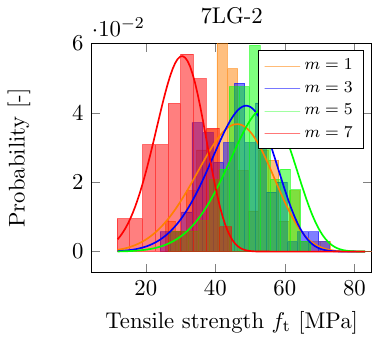}
    \caption{Layer-wise distribution of tensile strengths corresponding to the second most likely failure sequences $5 \rightarrow 1 + 3$~(sample 5LG) and $7 \rightarrow 1 + 3 + 5$~(samples 7LG-1 and 7LG-2). The layers are ordered in the top-to-bottom direction and enumerated by an index $m$ according to \figref{fig:domains} and the histograms are fitted with the Weibull distributions.}
    \label{fig:stoch_inverse}
\end{figure}

Nevertheless, we find it instructive to assess which strength combinations lead to the two-step cracking mechanisms by collecting the layer strength values for the realizations exhibiting the second most likely failure sequences in \figref{fig:stoch_inverse}. The data show that for all samples, the bottom layers exhibit the lowest strengths, followed by the top and interior layers. Such results support the fact that, in real-world applications, the outer layers have inferior mechanical properties to the interior ones because they are exposed to the surrounding environment and hence more prone to damage from transport and manipulation; see~\cite[Section~2]{bonati_probabilistic_2020} and \cite[Section 6.2]{alter2017enhanced} for further discussion. 

\section{Conclusions}\label{sec:conclusions}

This study used randomized phase field models to address simulations of distributed and gradual fracture patterns in laminated glass beams. As a building block, we used a layer-wise dimensionally-reduced phase-field model identified in our previous study~\cite{Schmidt:2020:PFF} as an efficient and accurate alternative to more refined two-dimensional plane strain or plate formulations. In particular, we adopted Timoshenko kinematics for each layer, ensured interlayer compatibility by means of an elimination technique, approximated the time- and the temperature-dependent response of an interlayer with a quasi-elastic model, and performed the numerical integration in the thickness direction to ensure that glass layer cross-sections were non-penetrable. Variability in the material properties was accounted for through a simple stochastic model, using the assumption that the strengths of individual layers are independent identical random variables with the Weibull distribution. Parameters of this distribution were determined from the data available in the literature and our own data for same-sized three-layer specimens.

From the performed combinatorial and Monte-Carlo simulations and experimental campaign for 5- and 7-layer glass beams, we concluded that:

\begin{enumerate}

    \item With a suitable combination of low- and high-strength layers, the model exhibits distributed cracking via an array of localized cracks.
    \item The second most likely failure patterns, which are in reasonable agreement with experimental results, correspond to lower strengths in the bottom and top layers, which are most exposed to the environment.  
    \item The single-layer benchmark proposed here can be used to test implementations of phase-field models for beam structures.
    \item The model fails to predict gradual cracking in individual layers in the bottom-up direction, as observed in our experiments. We attribute this limitation to neglecting strength variability along the length of the layers, which promotes the crack alignment in the adjacent layers.

\end{enumerate}

Extensions of this work could involve an extensive experimental campaign to obtain a representative set of structural responses, but this is beyond the scope of this paper. On the modeling side, a more refined description of strength variability would involve modeling strengths using random fields~\cite{Gerasimov2020,Hai2022,Nagaraja2023} and ensuring objectivity for the probability of failure under mesh refinement in the spirit of recent studies on crack band~\cite{Gorgogianni2022} and phase-field~\cite{hai_relationship_2023,wu_phase-field_2023} models. We plan to address this limitation in a future study, benefiting from the existing works on discrete element/peridynamics approaches to laminated glass fracture~\cite{biolzi_estimating_2017,casolo_modelling_2018} and insights from dedicated physically-based models for glass strength~\cite{symoens_probability_2023,rudshaug_physically_2023} and stochastic size effect considerations~\cite{bonati_redundancy_2019,bonati_probabilistic_2020}.

\section*{Author contributions}\noindent
J.~Schmidt: Conceptualization, Methodology, Software, Formal analysis, Data Curation, Writing - Original Draft, Visualization;
A.~Zemanová: Methodology, Formal analysis, Data Curation, Writing - Original Draft, Writing - Review \& Editing; 
J.~Zeman: Conceptualization, Methodology, Formal analysis, Writing - Original Draft, Writing - Review \& Editing

\section*{Acknowledgments}\noindent
We thank our colleagues from the Department of Steel and Timber Structures and the Experimental Center for providing us with data on the experimental campaign partially described in~\cite{Hana2018:FPB,Hana2020:FPB,konrad_laminated_2022} and Michal Šejnoha for helpful comments on the initial and revised versions of the manuscript. Besides, the anonymous referees are thanked for their critical suggestions for improving the manuscript quality. JS and AZ acknowledge the support from the Czech Science Foundation's project No.~22-15553S. JZ was supported by the European Regional Development Fund project Centre of Advanced Applied Sciences~(CAAS), No.~CZ 02.1.01/0.0/0.0/16\_019/0000778.

\appendix

\section{Derivatives of elastic energies}\label{app:energy_derivatives}

Because the polymer interlayers are assumed to be quasi-elastic, the energy derivative~\eqref{E:min_u_interlayer}, for $m = 2, 4, \ldots, M$, attains the standard linear form: 
\begin{align}\label{E:min_u_interlayer_grad}
    \bs{\nabla}_{\bs{u}_m}
    \Psi^\text{e}_{m}(t_n, \bs{u}_{m})
    \cdot
    \delta\bs{u}_{m}
    = & \,
    E_m 
    \int_0^L
    \Bigl( 
    A_m u_m^\prime \delta u_m^\prime 
    + 
    I_m
    \varphi^\prime_m \delta\varphi^\prime_m 
    \Bigr)
    \de x
    \\
    & +  
    G_m(t_n) A^*_m 
    \int_0^L
    \left( \varphi_m + w^\prime_m \right)
    \left( \delta\varphi_m + \delta w^\prime_m \right)
    \de x,
    \nonumber
\end{align}
For the glass layers, the explicit form of~\eqref{E:min_u_glass} for $m = 1, 3, \ldots, M$ is more involved:
\begin{align}\label{E:min_u_glass_full}
    \bs{\nabla}_{\bs{u}_m}    
    \Psi^\text{e}_{m}(\bs{u}_{m}, d_{m})
    = & 
    E_m \, b
    \int_0^L
    \sum_{j=1}^J
    \left( 
    (1 - d_m)^2
    \langle 
        u^\prime_m + \varphi^\prime_m z_{m,j}
    \rangle_+
    -
    \langle 
        u^\prime_m + \varphi^\prime_m z_{m,j}
    \rangle_-
    \right)
    \times \\
    & \left( \delta u_m' + \delta\varphi^\prime_m z_{m,j} \right)
    \Delta z_m
    \de x
    +
    G_m A_m^*
    \int_0^L
    (1 - d_m)^2 
    \left( \varphi_m + w_m^\prime \right)
    \left( \delta\varphi_m + \delta w_m^\prime \right)
    \de x,
    \nonumber
\end{align}
because of the numerical integration and the non-smooth non-linearity induced by the positive and negative parts $\langle \bullet \rangle_+$ and $\langle \bullet \rangle_-$.


\begin{thebibliography}{83}
\expandafter\ifx\csname natexlab\endcsname\relax\def\natexlab#1{#1}\fi
\providecommand{\url}[1]{\texttt{#1}}
\providecommand{\href}[2]{#2}
\providecommand{\path}[1]{#1}
\providecommand{\DOIprefix}{doi:}
\providecommand{\ArXivprefix}{arXiv:}
\providecommand{\URLprefix}{URL: }
\providecommand{\Pubmedprefix}{pmid:}
\providecommand{\doi}[1]{\href{http://dx.doi.org/#1}{\path{#1}}}
\providecommand{\Pubmed}[1]{\href{pmid:#1}{\path{#1}}}
\providecommand{\bibinfo}[2]{#2}
\ifx\xfnm\relax \def\xfnm[#1]{\unskip,\space#1}\fi
\bibitem[{Haldimann et~al.(2008)Haldimann, Luible, and
  Overend}]{Haldimann:2008:SUG}
\bibinfo{author}{M.~Haldimann}, \bibinfo{author}{A.~Luible},
  \bibinfo{author}{M.~Overend}, \bibinfo{title}{Structural Use of Glass},
  volume~\bibinfo{volume}{10} of \textit{\bibinfo{series}{Structural
  Engineering Documents}}, \bibinfo{publisher}{{IABSE}},
  \bibinfo{address}{Z\"{u}rich, Switzerland}, \bibinfo{year}{2008}.
\bibitem[{Bedon et~al.(2018)Bedon, Zhang, Santos, Honfi, Kozlowski, Arrigoni,
  Figuli, and Lange}]{Bedon2018_psf}
\bibinfo{author}{C.~Bedon}, \bibinfo{author}{X.~Zhang},
  \bibinfo{author}{F.~Santos}, \bibinfo{author}{D.~Honfi},
  \bibinfo{author}{M.~Kozlowski}, \bibinfo{author}{M.~Arrigoni},
  \bibinfo{author}{L.~Figuli}, \bibinfo{author}{D.~Lange},
\newblock \bibinfo{title}{Performance of structural glass facades under extreme
  loads -- {D}esign methods, existing research, current issues and trends},
\newblock \bibinfo{journal}{Construction and Building Materials}
  \bibinfo{volume}{163} (\bibinfo{year}{2018}) \bibinfo{pages}{921--937}.
  \DOIprefix\doi{10.1016/j.conbuildmat.2017.12.153}.
\bibitem[{Jozwik(2022)}]{Bonenberg2022}
\bibinfo{author}{A.~Jozwik},
\newblock \bibinfo{title}{Application of glass structures in architectural
  shaping of all-glass pavilions, extensions, and links},
\newblock \bibinfo{journal}{Buildings} \bibinfo{volume}{12}
  (\bibinfo{year}{2022}) \bibinfo{pages}{1254}.
  \DOIprefix\doi{10.3390/buildings12081254}.
\bibitem[{Galuppi and
  Royer-Carfagni(2012{\natexlab{a}})}]{galuppi2012effective}
\bibinfo{author}{L.~Galuppi}, \bibinfo{author}{G.~F. Royer-Carfagni},
\newblock \bibinfo{title}{Effective thickness of laminated glass beams: new
  expression via a variational approach},
\newblock \bibinfo{journal}{Engineering Structures} \bibinfo{volume}{38}
  (\bibinfo{year}{2012}{\natexlab{a}}) \bibinfo{pages}{53--67}.
  \DOIprefix\doi{10.1016/j.engstruct.2011.12.039}.
\bibitem[{Galuppi and
  Royer-Carfagni(2012{\natexlab{b}})}]{galuppi2012effectiveP}
\bibinfo{author}{L.~Galuppi}, \bibinfo{author}{G.~Royer-Carfagni},
\newblock \bibinfo{title}{The effective thickness of laminated glass plates},
\newblock \bibinfo{journal}{Journal of Mechanics of Materials and Structures}
  \bibinfo{volume}{7} (\bibinfo{year}{2012}{\natexlab{b}})
  \bibinfo{pages}{375--400}. \DOIprefix\doi{10.2140/jomms.2012.7.375}.
\bibitem[{López-Aenlle et~al.(2016)López-Aenlle, Pelayo, Ismael, Prieto,
  Martín~Rodríguez, and Fernández-Canteli}]{Lopez-Aenlle201644}
\bibinfo{author}{M.~López-Aenlle}, \bibinfo{author}{F.~Pelayo},
  \bibinfo{author}{G.~Ismael}, \bibinfo{author}{M.~Prieto},
  \bibinfo{author}{A.~Martín~Rodríguez},
  \bibinfo{author}{A.~Fernández-Canteli},
\newblock \bibinfo{title}{Buckling of laminated-glass beams using the
  effective-thickness concept},
\newblock \bibinfo{journal}{Composite Structures} \bibinfo{volume}{137}
  (\bibinfo{year}{2016}) \bibinfo{pages}{44--55}.
  \DOIprefix\doi{10.1016/j.compstruct.2015.11.014}.
\bibitem[{D'Ambrosio and Galuppi(2020)}]{D_Ambrosio2020205}
\bibinfo{author}{G.~D'Ambrosio}, \bibinfo{author}{L.~Galuppi},
\newblock \bibinfo{title}{Enhanced effective thickness model for buckling of
  {LG} beams with different boundary conditions},
\newblock \bibinfo{journal}{Glass Structures and Engineering}
  \bibinfo{volume}{5} (\bibinfo{year}{2020}) \bibinfo{pages}{205--210}.
  \DOIprefix\doi{10.1007/s40940-019-00116-3}.
\bibitem[{Aenlle and Pelayo(2013)}]{Aenlle2013}
\bibinfo{author}{M.~Aenlle}, \bibinfo{author}{F.~Pelayo},
\newblock \bibinfo{title}{Frequency response of laminated glass elements:
  {A}nalytical modeling and effective thickness},
\newblock \bibinfo{journal}{Applied Mechanics Reviews} \bibinfo{volume}{65}
  (\bibinfo{year}{2013}) \bibinfo{pages}{020802}.
  \DOIprefix\doi{10.1115/1.4023929}.
\bibitem[{Zemanov{\'a} et~al.(2018)Zemanov{\'a}, Zeman, Janda, Schmidt, and
  {\v{S}}ejnoha}]{zemanova2018modal}
\bibinfo{author}{A.~Zemanov{\'a}}, \bibinfo{author}{J.~Zeman},
  \bibinfo{author}{T.~Janda}, \bibinfo{author}{J.~Schmidt},
  \bibinfo{author}{M.~{\v{S}}ejnoha},
\newblock \bibinfo{title}{On modal analysis of laminated glass: {U}sability of
  simplified methods and enhanced effective thickness},
\newblock \bibinfo{journal}{Composites Part B: Engineering}
  \bibinfo{volume}{151} (\bibinfo{year}{2018}) \bibinfo{pages}{92--105}.
  \DOIprefix\doi{10.1016/j.compositesb.2018.05.032}.
\bibitem[{Foraboschi(2007)}]{foraboschi_behavior_2007}
\bibinfo{author}{P.~Foraboschi},
\newblock \bibinfo{title}{Behavior and failure strength of laminated glass
  beams},
\newblock \bibinfo{journal}{Journal of Engineering Mechanics}
  \bibinfo{volume}{133} (\bibinfo{year}{2007}) \bibinfo{pages}{1290--1301}.
  \DOIprefix\doi{10.1061/(ASCE)0733-9399(2007)133:12(1290)}.
\bibitem[{Foraboschi(2014)}]{foraboschi_optimal_2014}
\bibinfo{author}{P.~Foraboschi},
\newblock \bibinfo{title}{Optimal design of glass plates loaded transversally},
\newblock \bibinfo{journal}{Materials \& Design} \bibinfo{volume}{62}
  (\bibinfo{year}{2014}) \bibinfo{pages}{443--458}.
  \DOIprefix\doi{10.1016/j.matdes.2014.05.030}.
\bibitem[{Galuppi and Royer-Carfagni(2016)}]{galuppi_homogenized_2016}
\bibinfo{author}{L.~Galuppi}, \bibinfo{author}{G.~Royer-Carfagni},
\newblock \bibinfo{title}{A homogenized model for the post-breakage tensile
  behavior of laminated glass},
\newblock \bibinfo{journal}{Composite Structures} \bibinfo{volume}{154}
  (\bibinfo{year}{2016}) \bibinfo{pages}{600--615}.
  \DOIprefix\doi{10.1016/j.compstruct.2016.07.052}.
\bibitem[{Galuppi and Royer-Carfagni(2018)}]{galuppi_post-breakage_2018}
\bibinfo{author}{L.~Galuppi}, \bibinfo{author}{G.~Royer-Carfagni},
\newblock \bibinfo{title}{The post-breakage response of laminated heat-treated
  glass under in plane and out of plane loading},
\newblock \bibinfo{journal}{Composites Part B: Engineering}
  \bibinfo{volume}{147} (\bibinfo{year}{2018}) \bibinfo{pages}{227--239}.
  \DOIprefix\doi{10.1016/j.compositesb.2018.04.005}.
\bibitem[{Zang et~al.(2007)Zang, Lei, and Wang}]{Zang200773}
\bibinfo{author}{M.~Zang}, \bibinfo{author}{Z.~Lei}, \bibinfo{author}{S.~Wang},
\newblock \bibinfo{title}{Investigation of impact fracture behavior of
  automobile laminated glass by 3d discrete element method},
\newblock \bibinfo{journal}{Computational Mechanics} \bibinfo{volume}{41}
  (\bibinfo{year}{2007}) \bibinfo{pages}{73--83}.
  \DOIprefix\doi{10.1007/s00466-007-0170-1}.
\bibitem[{Baraldi et~al.(2016)Baraldi, Cecchi, and Foraboschi}]{Baraldi2016278}
\bibinfo{author}{D.~Baraldi}, \bibinfo{author}{A.~Cecchi},
  \bibinfo{author}{P.~Foraboschi},
\newblock \bibinfo{title}{Broken tempered laminated glass: {N}on-linear
  discrete element modeling},
\newblock \bibinfo{journal}{Composite Structures} \bibinfo{volume}{140}
  (\bibinfo{year}{2016}) \bibinfo{pages}{278--295}.
  \DOIprefix\doi{10.1016/j.compstruct.2015.12.050}.
\bibitem[{Chen et~al.(2016)Chen, Zang, Wang, Zheng, and Zhao}]{Chen20161}
\bibinfo{author}{S.~Chen}, \bibinfo{author}{M.~Zang},
  \bibinfo{author}{D.~Wang}, \bibinfo{author}{Z.~Zheng},
  \bibinfo{author}{C.~Zhao},
\newblock \bibinfo{title}{Finite element modelling of impact damage in
  polyvinyl butyral laminated glass},
\newblock \bibinfo{journal}{Composite Structures} \bibinfo{volume}{138}
  (\bibinfo{year}{2016}) \bibinfo{pages}{1--11}.
  \DOIprefix\doi{10.1016/j.compstruct.2015.11.042}.
\bibitem[{Vocialta et~al.(2018)Vocialta, Corrado, and
  Molinari}]{Vocialta2018448}
\bibinfo{author}{M.~Vocialta}, \bibinfo{author}{M.~Corrado},
  \bibinfo{author}{J.-F. Molinari},
\newblock \bibinfo{title}{Numerical analysis of fragmentation in tempered glass
  with parallel dynamic insertion of cohesive elements},
\newblock \bibinfo{journal}{Engineering Fracture Mechanics}
  \bibinfo{volume}{188} (\bibinfo{year}{2018}) \bibinfo{pages}{448--469}.
  \DOIprefix\doi{10.1016/j.engfracmech.2017.09.015}.
\bibitem[{Wu et~al.(2020)Wu, Wang, Huang, and Xu}]{Wu:2020:OSB}
\bibinfo{author}{L.~Wu}, \bibinfo{author}{L.~Wang}, \bibinfo{author}{D.~Huang},
  \bibinfo{author}{Y.~Xu},
\newblock \bibinfo{title}{An ordinary state-based peridynamic modeling for
  dynamic fracture of laminated glass under low-velocity impact},
\newblock \bibinfo{journal}{Composite Structures} \bibinfo{volume}{234}
  (\bibinfo{year}{2020}) \bibinfo{pages}{111722}.
  \DOIprefix\doi{10.1016/j.compstruct.2019.111722}.
\bibitem[{Naumenko et~al.(2022)Naumenko, Pander, and
  W\"{u}rkner}]{Naumenko2022}
\bibinfo{author}{K.~Naumenko}, \bibinfo{author}{M.~Pander},
  \bibinfo{author}{M.~W\"{u}rkner},
\newblock \bibinfo{title}{Damage patterns in float glass plates: {E}xperiments
  and peridynamics analysis},
\newblock \bibinfo{journal}{Theoretical and Applied Fracture Mechanics}
  \bibinfo{volume}{118} (\bibinfo{year}{2022}) \bibinfo{pages}{103264}.
  \DOIprefix\doi{10.1016/j.tafmec.2022.103264}.
\bibitem[{Wang et~al.(2017)Wang, Yang, Liu, Zhang, and Zhao}]{Wang2017493}
\bibinfo{author}{X.-E. Wang}, \bibinfo{author}{J.~Yang}, \bibinfo{author}{Q.-F.
  Liu}, \bibinfo{author}{Y.-M. Zhang}, \bibinfo{author}{C.~Zhao},
\newblock \bibinfo{title}{A comparative study of numerical modelling techniques
  for the fracture of brittle materials with specific reference to glass},
\newblock \bibinfo{journal}{Engineering Structures} \bibinfo{volume}{152}
  (\bibinfo{year}{2017}) \bibinfo{pages}{493--505}.
  \DOIprefix\doi{10.1016/j.engstruct.2017.08.050}.
\bibitem[{Teotia and Soni(2018)}]{Teotia2018412}
\bibinfo{author}{M.~Teotia}, \bibinfo{author}{R.~Soni},
\newblock \bibinfo{title}{Applications of finite element modelling in failure
  analysis of laminated glass composites: {A} review},
\newblock \bibinfo{journal}{Engineering Failure Analysis} \bibinfo{volume}{94}
  (\bibinfo{year}{2018}) \bibinfo{pages}{412--437}.
  \DOIprefix\doi{10.1016/j.engfailanal.2018.08.016}.
\bibitem[{Kuntsche et~al.(2019)Kuntsche, Schuster, and
  Schneider}]{Kuntsche2019209}
\bibinfo{author}{J.~Kuntsche}, \bibinfo{author}{M.~Schuster},
  \bibinfo{author}{J.~Schneider},
\newblock \bibinfo{title}{Engineering design of laminated safety glass
  considering the shear coupling: a review},
\newblock \bibinfo{journal}{Glass Structures and Engineering}
  \bibinfo{volume}{4} (\bibinfo{year}{2019}) \bibinfo{pages}{209--228}.
  \DOIprefix\doi{10.1007/s40940-019-00097-3}.
\bibitem[{Bourdin et~al.(2000)Bourdin, Francfort, and
  Marigo}]{bourdin2000numerical}
\bibinfo{author}{B.~Bourdin}, \bibinfo{author}{G.~A. Francfort},
  \bibinfo{author}{J.-J. Marigo},
\newblock \bibinfo{title}{Numerical experiments in revisited brittle fracture},
\newblock \bibinfo{journal}{Journal of the Mechanics and Physics of Solids}
  \bibinfo{volume}{48} (\bibinfo{year}{2000}) \bibinfo{pages}{797--826}.
  \DOIprefix\doi{10.1016/S0022-5096(99)00028-9}.
\bibitem[{Francfort and Marigo(1998)}]{Francfort19981319}
\bibinfo{author}{G.~Francfort}, \bibinfo{author}{J.-J. Marigo},
\newblock \bibinfo{title}{Revisiting brittle fracture as an energy minimization
  problem},
\newblock \bibinfo{journal}{Journal of the Mechanics and Physics of Solids}
  \bibinfo{volume}{46} (\bibinfo{year}{1998}) \bibinfo{pages}{1319--1342}.
  \DOIprefix\doi{10.1016/S0022-5096(98)00034-9}.
\bibitem[{Griffith(1921)}]{griffith1921vi}
\bibinfo{author}{A.~A. Griffith},
\newblock \bibinfo{title}{The phenomena of rupture and flow in solids},
\newblock \bibinfo{journal}{Philosophical Transactions of the Royal Society of
  London A} \bibinfo{volume}{221} (\bibinfo{year}{1921})
  \bibinfo{pages}{163--198}. \DOIprefix\doi{10.1098/rsta.1921.0006}.
\bibitem[{Bourdin et~al.(2008)Bourdin, Francfort, and
  Marigo}]{bourdin2008variational}
\bibinfo{author}{B.~Bourdin}, \bibinfo{author}{G.~A. Francfort},
  \bibinfo{author}{J.-J. Marigo},
\newblock \bibinfo{title}{The variational approach to fracture},
\newblock \bibinfo{journal}{Journal of Elasticity} \bibinfo{volume}{91}
  (\bibinfo{year}{2008}) \bibinfo{pages}{5--148}.
  \DOIprefix\doi{10.1007/s10659-007-9107-3}.
\bibitem[{Ambati et~al.(2015)Ambati, Gerasimov, and
  De~Lorenzis}]{ambati2015review}
\bibinfo{author}{M.~Ambati}, \bibinfo{author}{T.~Gerasimov},
  \bibinfo{author}{L.~De~Lorenzis},
\newblock \bibinfo{title}{A review on phase-field models of brittle fracture
  and a new fast hybrid formulation},
\newblock \bibinfo{journal}{Computational Mechanics} \bibinfo{volume}{55}
  (\bibinfo{year}{2015}) \bibinfo{pages}{383--405}.
  \DOIprefix\doi{10.1007/s00466-014-1109-y}.
\bibitem[{Wu et~al.(2020)Wu, Nguyen, Nguyen, Sutula, Sinaie, and
  Bordas}]{Wu2020}
\bibinfo{author}{J.~Y. Wu}, \bibinfo{author}{V.~P. Nguyen},
  \bibinfo{author}{C.~T. Nguyen}, \bibinfo{author}{D.~Sutula},
  \bibinfo{author}{S.~Sinaie}, \bibinfo{author}{S.~P. Bordas},
\newblock \bibinfo{title}{Phase-field modeling of fracture},
\newblock \bibinfo{journal}{Advances in Applied Mechanics} \bibinfo{volume}{53}
  (\bibinfo{year}{2020}) \bibinfo{pages}{1--183}.
  \DOIprefix\doi{10.1016/bs.aams.2019.08.001}.
\bibitem[{Amiri et~al.(2014)Amiri, Mill{\'{a}}n, Shen, Rabczuk, and
  Arroyo}]{Amiri2014}
\bibinfo{author}{F.~Amiri}, \bibinfo{author}{D.~Mill{\'{a}}n},
  \bibinfo{author}{Y.~Shen}, \bibinfo{author}{T.~Rabczuk},
  \bibinfo{author}{M.~Arroyo},
\newblock \bibinfo{title}{Phase-field modeling of fracture in linear thin
  shells},
\newblock \bibinfo{journal}{Theoretical and Applied Fracture Mechanics}
  \bibinfo{volume}{69} (\bibinfo{year}{2014}) \bibinfo{pages}{102--109}.
  \DOIprefix\doi{10.1016/j.tafmec.2013.12.002}.
\bibitem[{Kiendl et~al.(2016)Kiendl, Ambati, {De Lorenzis}, Gomez, and
  Reali}]{Kiendl2016}
\bibinfo{author}{J.~Kiendl}, \bibinfo{author}{M.~Ambati},
  \bibinfo{author}{L.~{De Lorenzis}}, \bibinfo{author}{H.~Gomez},
  \bibinfo{author}{A.~Reali},
\newblock \bibinfo{title}{Phase-field description of brittle fracture in plates
  and shells},
\newblock \bibinfo{journal}{Computer Methods in Applied Mechanics and
  Engineering} \bibinfo{volume}{312} (\bibinfo{year}{2016})
  \bibinfo{pages}{374--394}. \DOIprefix\doi{10.1016/j.cma.2016.09.011}.
\bibitem[{Kikis et~al.(2021)Kikis, Ambati, {De Lorenzis}, and
  Klinkel}]{Kikis2021}
\bibinfo{author}{G.~Kikis}, \bibinfo{author}{M.~Ambati},
  \bibinfo{author}{L.~{De Lorenzis}}, \bibinfo{author}{S.~Klinkel},
\newblock \bibinfo{title}{Phase-field model of brittle fracture in
  {Reissner}--{Mindlin} plates and shells},
\newblock \bibinfo{journal}{Computer Methods in Applied Mechanics and
  Engineering} \bibinfo{volume}{373} (\bibinfo{year}{2021})
  \bibinfo{pages}{113490}. \DOIprefix\doi{10.1016/j.cma.2020.113490}.
\bibitem[{Lai et~al.(2020)Lai, Gao, Li, Arroyo, and Shen}]{Lai2020}
\bibinfo{author}{W.~Lai}, \bibinfo{author}{J.~Gao}, \bibinfo{author}{Y.~Li},
  \bibinfo{author}{M.~Arroyo}, \bibinfo{author}{Y.~Shen},
\newblock \bibinfo{title}{Phase field modeling of brittle fracture in an
  {Euler}--{Bernoulli} beam accounting for transverse part-through cracks},
\newblock \bibinfo{journal}{Computer Methods in Applied Mechanics and
  Engineering} \bibinfo{volume}{361} (\bibinfo{year}{2020})
  \bibinfo{pages}{112787}. \DOIprefix\doi{10.1016/j.cma.2019.112787}.
\bibitem[{Ambati et~al.(2022)Ambati, Heinzmann, Seiler, and
  K{\"{a}}stner}]{Ambati2022}
\bibinfo{author}{M.~Ambati}, \bibinfo{author}{J.~Heinzmann},
  \bibinfo{author}{M.~Seiler}, \bibinfo{author}{M.~K{\"{a}}stner},
\newblock \bibinfo{title}{Phase-field modeling of brittle fracture along the
  thickness direction of plates and shells},
\newblock \bibinfo{journal}{International Journal for Numerical Methods in
  Engineering} \bibinfo{volume}{123} (\bibinfo{year}{2022})
  \bibinfo{pages}{4094--4118}. \DOIprefix\doi{10.1002/nme.7001}.
\bibitem[{Bijaya et~al.(2023)Bijaya, {Roy Chowdhury}, and
  Chowdhury}]{Bijaya2023}
\bibinfo{author}{A.~Bijaya}, \bibinfo{author}{S.~{Roy Chowdhury}},
  \bibinfo{author}{R.~Chowdhury},
\newblock \bibinfo{title}{Reduced-dimensional phase-field theory for lattice
  fracture and its application in fracture toughness assessment of architected
  materials},
\newblock \bibinfo{journal}{European Journal of Mechanics - A/Solids}
  \bibinfo{volume}{100} (\bibinfo{year}{2023}) \bibinfo{pages}{104964}.
  \DOIprefix\doi{10.1016/j.euromechsol.2023.104964}.
\bibitem[{Alessi and Freddi(2017)}]{ALESSI20179}
\bibinfo{author}{R.~Alessi}, \bibinfo{author}{F.~Freddi},
\newblock \bibinfo{title}{Phase-field modelling of failure in hybrid
  laminates},
\newblock \bibinfo{journal}{Composite Structures} \bibinfo{volume}{181}
  (\bibinfo{year}{2017}) \bibinfo{pages}{9--25}.
  \DOIprefix\doi{10.1016/j.compstruct.2017.08.073}.
\bibitem[{Alessi and Freddi(2019)}]{Alessi2019}
\bibinfo{author}{R.~Alessi}, \bibinfo{author}{F.~Freddi},
\newblock \bibinfo{title}{Failure and complex crack patterns in hybrid
  laminates: {A} phase-field approach},
\newblock \bibinfo{journal}{Composites Part B: Engineering}
  (\bibinfo{year}{2019}) \bibinfo{pages}{107256}.
  \DOIprefix\doi{10.1016/j.compositesb.2019.107256}.
\bibitem[{Freddi and Mingazzi(2020)}]{Freddi2020}
\bibinfo{author}{F.~Freddi}, \bibinfo{author}{L.~Mingazzi},
\newblock \bibinfo{title}{Phase field simulation of laminated glass beam},
\newblock \bibinfo{journal}{Materials} \bibinfo{volume}{13}
  (\bibinfo{year}{2020}) \bibinfo{pages}{3218}.
  \DOIprefix\doi{10.3390/ma13143218}.
\bibitem[{Schmidt et~al.(2020)Schmidt, Zemanová, Zeman, and
  \v{S}ejnoha}]{Schmidt:2020:PFF}
\bibinfo{author}{J.~Schmidt}, \bibinfo{author}{A.~Zemanová},
  \bibinfo{author}{J.~Zeman}, \bibinfo{author}{M.~\v{S}ejnoha},
\newblock \bibinfo{title}{Phase-field fracture modelling of thin monolithic and
  laminated glass plates under quasi-static bending},
\newblock \bibinfo{journal}{Materials} \bibinfo{volume}{13}
  (\bibinfo{year}{2020}) \bibinfo{pages}{5153}.
  \DOIprefix\doi{10.3390/ma13225153}.
  \href{http://arxiv.org/abs/2010.00375}{{\tt arXiv:2010.00375}}.
\bibitem[{Schmidt et~al.(2023)Schmidt, Janda, and Šejnoha}]{Schmidt:2023:PPP}
\bibinfo{author}{J.~Schmidt}, \bibinfo{author}{T.~Janda},
  \bibinfo{author}{M.~Šejnoha},
\newblock \bibinfo{title}{Prediction of pre- and post-breakage behavior of
  laminated glass using a phase-field damage model},
\newblock \bibinfo{journal}{Applied Sciences} \bibinfo{volume}{13}
  (\bibinfo{year}{2023}) \bibinfo{pages}{1708}.
  \DOIprefix\doi{10.3390/app13031708}.
\bibitem[{Weibull(1951)}]{Weibull1951}
\bibinfo{author}{W.~Weibull},
\newblock \bibinfo{title}{A statistical distribution function of wide
  applicability},
\newblock \bibinfo{journal}{Journal of Applied Mechanics} \bibinfo{volume}{18}
  (\bibinfo{year}{1951}) \bibinfo{pages}{293--297}. \URLprefix
  \url{https://hal.archives-ouvertes.fr/hal-03112318}.
  \DOIprefix\doi{10.1115/1.4010337}.
\bibitem[{Bonati et~al.(2019)Bonati, Pisano, and
  Royer~Carfagni}]{bonati_redundancy_2019}
\bibinfo{author}{A.~Bonati}, \bibinfo{author}{G.~Pisano},
  \bibinfo{author}{G.~Royer~Carfagni},
\newblock \bibinfo{title}{Redundancy and robustness of brittle laminated
  plates. {Overlooked} aspects in structural glass},
\newblock \bibinfo{journal}{Composite Structures} \bibinfo{volume}{227}
  (\bibinfo{year}{2019}) \bibinfo{pages}{111288}.
  \DOIprefix\doi{10.1016/j.compstruct.2019.111288}.
\bibitem[{Bonati et~al.(2020)Bonati, Pisano, and
  Royer-Carfagni}]{bonati_probabilistic_2020}
\bibinfo{author}{A.~Bonati}, \bibinfo{author}{G.~Pisano},
  \bibinfo{author}{G.~Royer-Carfagni},
\newblock \bibinfo{title}{Probabilistic considerations about the strength of
  laminated annealed float glass},
\newblock \bibinfo{journal}{Glass Structures \& Engineering}
  \bibinfo{volume}{5} (\bibinfo{year}{2020}) \bibinfo{pages}{27--40}.
  \DOIprefix\doi{10.1007/s40940-019-00111-8}.
\bibitem[{Biolzi et~al.(2017)Biolzi, Casolo, Diana, and
  Sanjust}]{biolzi_estimating_2017}
\bibinfo{author}{L.~Biolzi}, \bibinfo{author}{S.~Casolo},
  \bibinfo{author}{V.~Diana}, \bibinfo{author}{C.~A. Sanjust},
\newblock \bibinfo{title}{Estimating laminated glass beam strength via
  stochastic rigid body-spring model},
\newblock \bibinfo{journal}{Composite Structures} \bibinfo{volume}{172}
  (\bibinfo{year}{2017}) \bibinfo{pages}{61--72}.
  \DOIprefix\doi{10.1016/j.compstruct.2017.03.062}.
\bibitem[{Casolo and Diana(2018)}]{casolo_modelling_2018}
\bibinfo{author}{S.~Casolo}, \bibinfo{author}{V.~Diana},
\newblock \bibinfo{title}{Modelling laminated glass beam failure via stochastic
  rigid body-spring model and bond-based peridynamics},
\newblock \bibinfo{journal}{Engineering Fracture Mechanics}
  \bibinfo{volume}{190} (\bibinfo{year}{2018}) \bibinfo{pages}{331--346}.
  \DOIprefix\doi{10.1016/j.engfracmech.2017.12.028}.
\bibitem[{Hai et~al.(2023)Hai, Li, and Wriggers}]{hai_relationship_2023}
\bibinfo{author}{L.~Hai}, \bibinfo{author}{J.~Li},
  \bibinfo{author}{P.~Wriggers},
\newblock \bibinfo{title}{Relationship between probabilistic characteristics of
  microscopic and macroscopic strength within the stochastic phase-field
  model},
\newblock \bibinfo{journal}{Applied Mathematical Modelling}
  \bibinfo{volume}{123} (\bibinfo{year}{2023}) \bibinfo{pages}{776--789}.
  \DOIprefix\doi{10.1016/j.apm.2023.07.027}.
\bibitem[{Wu et~al.(2023)Wu, Yao, and Le}]{wu_phase-field_2023}
\bibinfo{author}{J.-Y. Wu}, \bibinfo{author}{J.-R. Yao}, \bibinfo{author}{J.-L.
  Le},
\newblock \bibinfo{title}{Phase-field modeling of stochastic fracture in
  heterogeneous quasi-brittle solids},
\newblock \bibinfo{journal}{Computer Methods in Applied Mechanics and
  Engineering} \bibinfo{volume}{416} (\bibinfo{year}{2023})
  \bibinfo{pages}{116332}. \DOIprefix\doi{10.1016/j.cma.2023.116332}.
\bibitem[{Galuppi and Royer-Carfagni(2012)}]{Galuppi2012}
\bibinfo{author}{L.~Galuppi}, \bibinfo{author}{G.~Royer-Carfagni},
\newblock \bibinfo{title}{{Laminated beams with viscoelastic interlayer}},
\newblock \bibinfo{journal}{International Journal of Solids and Structures}
  \bibinfo{volume}{49} (\bibinfo{year}{2012}) \bibinfo{pages}{2637--2645}.
  \DOIprefix\doi{10.1016/J.ijsolstr.2012.05.028}.
\bibitem[{Galuppi and Royer-Carfagni(2013)}]{Galuppi2013}
\bibinfo{author}{L.~Galuppi}, \bibinfo{author}{G.~Royer-Carfagni},
\newblock \bibinfo{title}{{The design of laminated glass under time-dependent
  loading}},
\newblock \bibinfo{journal}{International Journal of Mechanical Sciences}
  \bibinfo{volume}{68} (\bibinfo{year}{2013}) \bibinfo{pages}{67--75}.
  \DOIprefix\doi{10.1016/j.ijmecsci.2012.12.019}.
\bibitem[{Zemanov\'{a} et~al.(2017)Zemanov\'{a}, Zeman, and
  \v{S}ejnoha}]{ZEMANOVA2017380}
\bibinfo{author}{A.~Zemanov\'{a}}, \bibinfo{author}{J.~Zeman},
  \bibinfo{author}{M.~\v{S}ejnoha},
\newblock \bibinfo{title}{Comparison of viscoelastic finite element models for
  laminated glass beams},
\newblock \bibinfo{journal}{International Journal of Mechanical Sciences}
  \bibinfo{volume}{131--132} (\bibinfo{year}{2017}) \bibinfo{pages}{380--395}.
  \DOIprefix\doi{10.1016/j.ijmecsci.2017.05.035}.
\bibitem[{Chen et~al.(2017)Chen, Zang, Wang, Yoshimura, and
  Yamada}]{chen2017numerical}
\bibinfo{author}{S.~Chen}, \bibinfo{author}{M.~Zang},
  \bibinfo{author}{D.~Wang}, \bibinfo{author}{S.~Yoshimura},
  \bibinfo{author}{T.~Yamada},
\newblock \bibinfo{title}{Numerical analysis of impact failure of automotive
  laminated glass: {A} review},
\newblock \bibinfo{journal}{Composites Part B: Engineering}
  \bibinfo{volume}{122} (\bibinfo{year}{2017}) \bibinfo{pages}{47--60}.
  \DOIprefix\doi{10.1016/j.compositesb.2017.04.007}.
\bibitem[{Timoshenko(1921)}]{timoshenko_correction_1921}
\bibinfo{author}{S.~Timoshenko},
\newblock \bibinfo{title}{On the correction for shear of the differential
  equation for transverse vibrations of prismatic bars},
\newblock \bibinfo{journal}{The London, Edinburgh, and Dublin Philosophical
  Magazine and Journal of Science} \bibinfo{volume}{41} (\bibinfo{year}{1921})
  \bibinfo{pages}{744--746}. \DOIprefix\doi{62610.1080/14786442108634}.
\bibitem[{Timoshenko(1922)}]{timoshenko_transverse_1922}
\bibinfo{author}{S.~Timoshenko},
\newblock \bibinfo{title}{On the transverse vibrations of bars of uniform
  cross-section},
\newblock \bibinfo{journal}{The London, Edinburgh, and Dublin Philosophical
  Magazine and Journal of Science} \bibinfo{volume}{43} (\bibinfo{year}{1922})
  \bibinfo{pages}{125--131}. \DOIprefix\doi{10.1080/14786442208633855}.
\bibitem[{Duser et~al.(1999)Duser, Jagota, and Bennison}]{Duser1999}
\bibinfo{author}{A.~V. Duser}, \bibinfo{author}{A.~Jagota},
  \bibinfo{author}{S.~J. Bennison},
\newblock \bibinfo{title}{Analysis of glass / {PolyVinyl Butyral} laminates
  subjected to uniform pressure},
\newblock \bibinfo{journal}{Journal of Engineering Mechanics}
  \bibinfo{volume}{125} (\bibinfo{year}{1999}) \bibinfo{pages}{435--442}.
  \DOIprefix\doi{10.1061/(ASCE)0733-9399(1999)125:4(435)}.
\bibitem[{H{\'a}na et~al.(2019)H{\'a}na, Janda, Schmidt, Zemanov{\'a},
  {\v{S}}ejnoha, Eli{\'a}{\v{s}}ov{\'a}, and
  Vok{\'a}{\v{c}}}]{hana2019experimental}
\bibinfo{author}{T.~H{\'a}na}, \bibinfo{author}{T.~Janda},
  \bibinfo{author}{J.~Schmidt}, \bibinfo{author}{A.~Zemanov{\'a}},
  \bibinfo{author}{M.~{\v{S}}ejnoha},
  \bibinfo{author}{M.~Eli{\'a}{\v{s}}ov{\'a}},
  \bibinfo{author}{M.~Vok{\'a}{\v{c}}},
\newblock \bibinfo{title}{Experimental and numerical study of viscoelastic
  properties of polymeric interlayers used for laminated glass: Determination
  of material parameters},
\newblock \bibinfo{journal}{Materials} \bibinfo{volume}{12}
  (\bibinfo{year}{2019}) \bibinfo{pages}{2241}.
  \DOIprefix\doi{10.3390/ma12142241}.
\bibitem[{Miehe et~al.(2010)Miehe, Welschinger, and
  Hofacker}]{miehe2010thermodynamically}
\bibinfo{author}{C.~Miehe}, \bibinfo{author}{F.~Welschinger},
  \bibinfo{author}{M.~Hofacker},
\newblock \bibinfo{title}{Thermodynamically consistent phase-field models of
  fracture: {V}ariational principles and multi-field {FE} implementations},
\newblock \bibinfo{journal}{International Journal for Numerical Methods in
  Engineering} \bibinfo{volume}{83} (\bibinfo{year}{2010})
  \bibinfo{pages}{1273--1311}. \DOIprefix\doi{10.1002/nme.2861}.
\bibitem[{Pham et~al.(2011)Pham, Amor, Marigo, and Maurini}]{pham2011gradient}
\bibinfo{author}{K.~Pham}, \bibinfo{author}{H.~Amor}, \bibinfo{author}{J.-J.
  Marigo}, \bibinfo{author}{C.~Maurini},
\newblock \bibinfo{title}{Gradient damage models and their use to approximate
  brittle fracture},
\newblock \bibinfo{journal}{International Journal of Damage Mechanics}
  \bibinfo{volume}{20} (\bibinfo{year}{2011}) \bibinfo{pages}{618--652}.
  \URLprefix \url{https://hal.archives-ouvertes.fr/hal-00549530}.
  \DOIprefix\doi{10.1177/1056789510386852}.
\bibitem[{Kiendl et~al.(2016)Kiendl, Ambati, De~Lorenzis, Gomez, and
  Reali}]{kiendl2016phase}
\bibinfo{author}{J.~Kiendl}, \bibinfo{author}{M.~Ambati},
  \bibinfo{author}{L.~De~Lorenzis}, \bibinfo{author}{H.~Gomez},
  \bibinfo{author}{A.~Reali},
\newblock \bibinfo{title}{Phase-field description of brittle fracture in plates
  and shells},
\newblock \bibinfo{journal}{Computer Methods in Applied Mechanics and
  Engineering} \bibinfo{volume}{312} (\bibinfo{year}{2016})
  \bibinfo{pages}{374--394}. \DOIprefix\doi{10.1016/j.cma.2016.09.011}.
\bibitem[{Jir\'{a}sek and Zeman(2015)}]{Jirasek:2015:LSR}
\bibinfo{author}{M.~Jir\'{a}sek}, \bibinfo{author}{J.~Zeman},
\newblock \bibinfo{title}{Localization study of a regularized variational
  damage model},
\newblock \bibinfo{journal}{International Journal of Solids and Structures}
  \bibinfo{volume}{69--70} (\bibinfo{year}{2015}) \bibinfo{pages}{131--151}.
  \DOIprefix\doi{10.1016/j.ijsolstr.2015.06.001}.
  \href{http://arxiv.org/abs/1412.5539}{{\tt arXiv:1412.5539}}.
\bibitem[{Miehe et~al.(2015)Miehe, Schaenzel, and Ulmer}]{miehe2015phase}
\bibinfo{author}{C.~Miehe}, \bibinfo{author}{L.-M. Schaenzel},
  \bibinfo{author}{H.~Ulmer},
\newblock \bibinfo{title}{Phase field modeling of fracture in multi-physics
  problems. {Part I}. balance of crack surface and failure criteria for brittle
  crack propagation in thermo-elastic solids},
\newblock \bibinfo{journal}{Computer Methods in Applied Mechanics and
  Engineering} \bibinfo{volume}{294} (\bibinfo{year}{2015})
  \bibinfo{pages}{449--485}. \DOIprefix\doi{10.1016/j.cma.2014.11.016}.
\bibitem[{Bilgen and Weinberg(2019)}]{Bilgen2019}
\bibinfo{author}{C.~Bilgen}, \bibinfo{author}{K.~Weinberg},
\newblock \bibinfo{title}{On the crack-driving force of phase-field models in
  linearized and finite elasticity},
\newblock \bibinfo{journal}{Computer Methods in Applied Mechanics and
  Engineering} \bibinfo{volume}{353} (\bibinfo{year}{2019})
  \bibinfo{pages}{348--372}. \DOIprefix\doi{10.1016/j.cma.2019.05.009}.
  \href{http://arxiv.org/abs/1808.00542}{{\tt arXiv:1808.00542}}.
\bibitem[{Logg et~al.(2012)Logg, Mardal, and Wells}]{logg2012automated}
\bibinfo{author}{A.~Logg}, \bibinfo{author}{K.-A. Mardal},
  \bibinfo{author}{G.~Wells}, \bibinfo{title}{Automated solution of
  differential equations by the finite element method: The FEniCS book},
  volume~\bibinfo{volume}{84}, \bibinfo{publisher}{Springer Science \& Business
  Media}, \bibinfo{year}{2012}. \DOIprefix\doi{10.1007/978-3-642-23099-8}.
\bibitem[{Schmidt(2023)}]{git_article}
\bibinfo{author}{J.~Schmidt}, \bibinfo{title}{Phase-field analysis of laminated
  glass - supplementary code for the article}, \bibinfo{year}{2023}. \URLprefix
  \url{https://gitlab.com/js_workdir/article_support/stoch_pf_beams},
  \bibinfo{note}{[accessed 23 October 2023]}.
\bibitem[{Dujc and Brank(2022)}]{Dujc2021}
\bibinfo{author}{J.~Dujc}, \bibinfo{author}{B.~Brank},
\newblock \bibinfo{title}{Combining coupled, staggered and uncoupled solution
  methods for phase-field-based fracture analysis},
\newblock \bibinfo{journal}{Mechanics of Advanced Materials and Structures}
  \bibinfo{volume}{29} (\bibinfo{year}{2022}) \bibinfo{pages}{6361--6378}.
  \DOIprefix\doi{10.1080/15376494.2021.1976888}.
\bibitem[{{European Committee for Standardization}(2000)}]{EN1288-3}
\bibinfo{author}{{European Committee for Standardization}}, \bibinfo{title}{{EN
  1288-3:2000 Glass in building -- Determination of the bending strength of
  glass -- Part 3: Test with specimen supported at two points (four point
  bending)}}, \bibinfo{year}{2000}.
\bibitem[{Hána et~al.(2018)Hána, Eliášová, and Sokol}]{Hana2018:FPB}
\bibinfo{author}{T.~Hána}, \bibinfo{author}{M.~Eliášová},
  \bibinfo{author}{Z.~Sokol},
\newblock \bibinfo{title}{Four point bending tests of double laminated glass
  panels},
\newblock in: \bibinfo{editor}{F.~C.}, \bibinfo{editor}{J.~Náprstek} (Eds.),
  \bibinfo{booktitle}{Engineering Mechanics 2018: Book of Full Texts},
  \bibinfo{publisher}{Institute of Theoretical and Applied Mechanics, Czech
  Academy of Sciences}, \bibinfo{address}{Svratka, Czech Republic},
  \bibinfo{year}{2018}, pp. \bibinfo{pages}{285--288}.
  \DOIprefix\doi{10.21495/91-8-285}.
\bibitem[{Hána et~al.(2020)Hána, Vokáč, Eliášová, Sokol, and
  Machalická-Vokáčová}]{Hana2020:FPB}
\bibinfo{author}{T.~Hána}, \bibinfo{author}{M.~Vokáč},
  \bibinfo{author}{M.~Eliášová}, \bibinfo{author}{Z.~Sokol},
  \bibinfo{author}{K.~Machalická-Vokáčová},
\newblock \bibinfo{title}{Four-point bending tests of {PVB} double laminated
  glass panels -- {Experiments} and numerical analysis},
\newblock in: \bibinfo{editor}{J.~Belis}, \bibinfo{editor}{F.~Bos},
  \bibinfo{editor}{C.~Louter} (Eds.), \bibinfo{booktitle}{Challenging Glass 7:
  Conference on Architectural and Structural Applications of Glass},
  \bibinfo{publisher}{Delft University of Technology}, \bibinfo{address}{Ghent
  University, Belgium}, \bibinfo{year}{2020}, p.~\bibinfo{pages}{12}.
  \DOIprefix\doi{10.7480/cgc.7.4460}.
\bibitem[{Konrád et~al.(2022)Konrád, Hála, Schmidt, Zemanová, and
  Sovják}]{konrad_laminated_2022}
\bibinfo{author}{P.~Konrád}, \bibinfo{author}{P.~Hála},
  \bibinfo{author}{J.~Schmidt}, \bibinfo{author}{A.~Zemanová},
  \bibinfo{author}{R.~Sovják},
\newblock \bibinfo{title}{Laminated glass plates subjected to high-velocity
  projectile impact and their residual post-impact performance},
\newblock \bibinfo{journal}{Materials} \bibinfo{volume}{15}
  (\bibinfo{year}{2022}) \bibinfo{pages}{8342}.
  \DOIprefix\doi{10.3390/ma15238342}.
\bibitem[{Haldimann et~al.(2008)Haldimann, Luible, and
  Overend}]{haldimann2008structural}
\bibinfo{author}{M.~Haldimann}, \bibinfo{author}{A.~Luible},
  \bibinfo{author}{M.~Overend}, \bibinfo{title}{Structural use of glass},
  volume~\bibinfo{volume}{10}, \bibinfo{publisher}{Iabse},
  \bibinfo{year}{2008}.
\bibitem[{Veer et~al.(2009)Veer, Louter, and Bos}]{veer2009}
\bibinfo{author}{F.~A. Veer}, \bibinfo{author}{P.~C. Louter},
  \bibinfo{author}{F.~P. Bos},
\newblock \bibinfo{title}{The strength of annealed, heat-strengthened and fully
  tempered float glass},
\newblock \bibinfo{journal}{Fatigue \& Fracture of Engineering Materials \&
  Structures} \bibinfo{volume}{32} (\bibinfo{year}{2009})
  \bibinfo{pages}{18--25}. \DOIprefix\doi{10.1111/j.1460-2695.2008.01308.x}.
\bibitem[{Zemanová et~al.(2018)Zemanová, Schmidt, and
  \v{S}ejnoha}]{Zemanova2018}
\bibinfo{author}{A.~Zemanová}, \bibinfo{author}{J.~Schmidt},
  \bibinfo{author}{M.~\v{S}ejnoha},
\newblock \bibinfo{title}{Evaluation of tensile strength of glass from combined
  experimental and numerical analysis of laminated glass},
\newblock in: \bibinfo{booktitle}{WIT Transactions on the Built Environment},
  volume \bibinfo{volume}{175}, \bibinfo{year}{2018}, pp.
  \bibinfo{pages}{29--39}. \DOIprefix\doi{10.2495/HPSM180041}.
\bibitem[{Castori and Speranzini(2019)}]{Castori2019}
\bibinfo{author}{G.~Castori}, \bibinfo{author}{E.~Speranzini},
\newblock \bibinfo{title}{Fracture strength prediction of float glass: The
  coaxial double ring test method},
\newblock \bibinfo{journal}{Construction and Building Materials}
  \bibinfo{volume}{225} (\bibinfo{year}{2019}) \bibinfo{pages}{1064--1076}.
  \DOIprefix\doi{10.1016/j.conbuildmat.2019.07.264}.
\bibitem[{Müller-Braun et~al.(2020)Müller-Braun, Seel, König, Hof,
  Schneider, and Oechsner}]{braun2020cut}
\bibinfo{author}{S.~Müller-Braun}, \bibinfo{author}{M.~Seel},
  \bibinfo{author}{M.~König}, \bibinfo{author}{P.~Hof},
  \bibinfo{author}{J.~Schneider}, \bibinfo{author}{M.~Oechsner},
\newblock \bibinfo{title}{Cut edge of annealed float glass: crack system and
  possibilities to increase the edge strength by adjusting the cutting
  process},
\newblock \bibinfo{journal}{Glass Structures and Engineering}
  \bibinfo{volume}{5} (\bibinfo{year}{2020}) \bibinfo{pages}{3--25}.
  \DOIprefix\doi{10.1007/S40940-019-00108-3}.
\bibitem[{Bukieda et~al.(2020)Bukieda, Lohr, Meiberg, and Weller}]{Bukieda2020}
\bibinfo{author}{P.~Bukieda}, \bibinfo{author}{K.~Lohr},
  \bibinfo{author}{J.~Meiberg}, \bibinfo{author}{B.~Weller},
\newblock \bibinfo{title}{Study on the optical quality and strength of glass
  edges after the grinding and polishing process},
\newblock \bibinfo{journal}{Glass Structures and Engineering}
  \bibinfo{volume}{5} (\bibinfo{year}{2020}) \bibinfo{pages}{411--428}.
  \DOIprefix\doi{10.1007/S40940-020-00121-X}.
\bibitem[{Bažant et~al.(2021)Bažant, Le, and Salviato}]{Bazant:2021}
\bibinfo{author}{Z.~P. Bažant}, \bibinfo{author}{J.-L. Le},
  \bibinfo{author}{M.~Salviato}, \bibinfo{title}{Quasibrittle fracture
  mechanics and size effect: A first course}, \bibinfo{publisher}{Oxford
  University Press}, \bibinfo{year}{2021}.
  \DOIprefix\doi{10.1093/oso/9780192846242.001.0001}.
\bibitem[{Wu and Nguyen(2018)}]{wu2018length}
\bibinfo{author}{J.-Y. Wu}, \bibinfo{author}{V.~P. Nguyen},
\newblock \bibinfo{title}{A length scale insensitive phase-field damage model
  for brittle fracture},
\newblock \bibinfo{journal}{Journal of the Mechanics and Physics of Solids}
  \bibinfo{volume}{119} (\bibinfo{year}{2018}) \bibinfo{pages}{20--42}.
  \DOIprefix\doi{10.1016/j.jmps.2018.06.006}.
\bibitem[{Gross et~al.(2018)Gross, Hauger, Schröder, Wall, and
  Bonet}]{Gross:2018:EM2}
\bibinfo{author}{D.~Gross}, \bibinfo{author}{W.~Hauger},
  \bibinfo{author}{J.~Schröder}, \bibinfo{author}{W.~A. Wall},
  \bibinfo{author}{J.~Bonet}, \bibinfo{title}{Engineering mechanics 2:
  Mechanics of materials}, \bibinfo{edition}{second} ed.,
  \bibinfo{publisher}{Springer-Verlag GmbH}, \bibinfo{year}{2018}.
  \DOIprefix\doi{10.1007/978-3-662-56272-7}.
\bibitem[{Alter et~al.(2017)Alter, Kolling, and Schneider}]{alter2017enhanced}
\bibinfo{author}{C.~Alter}, \bibinfo{author}{S.~Kolling},
  \bibinfo{author}{J.~Schneider},
\newblock \bibinfo{title}{An enhanced non--local failure criterion for
  laminated glass under low velocity impact},
\newblock \bibinfo{journal}{International Journal of Impact Engineering}
  \bibinfo{volume}{109} (\bibinfo{year}{2017}) \bibinfo{pages}{342--353}.
\bibitem[{Gerasimov et~al.(2020)Gerasimov, R{\"{o}}mer, Vondřejc, Matthies,
  and {de Lorenzis}}]{Gerasimov2020}
\bibinfo{author}{T.~Gerasimov}, \bibinfo{author}{U.~R{\"{o}}mer},
  \bibinfo{author}{J.~Vondřejc}, \bibinfo{author}{H.~G. Matthies},
  \bibinfo{author}{L.~{de Lorenzis}},
\newblock \bibinfo{title}{Stochastic phase-field modeling of brittle fracture:
  {Computing} multiple crack patterns and their probabilities},
\newblock \bibinfo{journal}{Computer Methods in Applied Mechanics and
  Engineering} \bibinfo{volume}{372} (\bibinfo{year}{2020})
  \bibinfo{pages}{113353}. \DOIprefix\doi{10.1016/J.cma.2020.113353}.
  \href{http://arxiv.org/abs/2005.01332}{{\tt arXiv:2005.01332}}.
\bibitem[{Hai and Li(2022)}]{Hai2022}
\bibinfo{author}{L.~Hai}, \bibinfo{author}{J.~Li},
\newblock \bibinfo{title}{Modeling tensile damage and fracture of quasi-brittle
  materials using stochastic phase-field model},
\newblock \bibinfo{journal}{Theoretical and Applied Fracture Mechanics}
  \bibinfo{volume}{118} (\bibinfo{year}{2022}) \bibinfo{pages}{103283}.
  \DOIprefix\doi{10.1016/j.tafmec.2022.103283}.
\bibitem[{Nagaraja et~al.(2023)Nagaraja, R{\"{o}}mer, Matthies, and {de
  Lorenzis}}]{Nagaraja2023}
\bibinfo{author}{S.~Nagaraja}, \bibinfo{author}{U.~R{\"{o}}mer},
  \bibinfo{author}{H.~G. Matthies}, \bibinfo{author}{L.~{de Lorenzis}},
\newblock \bibinfo{title}{{Deterministic and stochastic phase-field modeling of
  anisotropic brittle fracture}},
\newblock \bibinfo{journal}{Computer Methods in Applied Mechanics and
  Engineering} \bibinfo{volume}{408} (\bibinfo{year}{2023})
  \bibinfo{pages}{115960}. \DOIprefix\doi{10.1016/j.cma.2023.115960}.
\bibitem[{Gorgogianni et~al.(2022)Gorgogianni, Eli{\'{a}}{\v{s}}, and
  Le}]{Gorgogianni2022}
\bibinfo{author}{A.~Gorgogianni}, \bibinfo{author}{J.~Eli{\'{a}}{\v{s}}},
  \bibinfo{author}{J.~L. Le},
\newblock \bibinfo{title}{Mesh objective stochastic simulations of quasibrittle
  fracture},
\newblock \bibinfo{journal}{Journal of the Mechanics and Physics of Solids}
  \bibinfo{volume}{159} (\bibinfo{year}{2022}) \bibinfo{pages}{104745}.
  \DOIprefix\doi{10.1016/j.jmps.2021.104745}.
\bibitem[{Symoens et~al.(2023)Symoens, Van~Coile, Jovanović, and
  Belis}]{symoens_probability_2023}
\bibinfo{author}{E.~Symoens}, \bibinfo{author}{R.~Van~Coile},
  \bibinfo{author}{B.~Jovanović}, \bibinfo{author}{J.~Belis},
\newblock \bibinfo{title}{Probability density function models for float glass
  under mechanical loading with varying parameters},
\newblock \bibinfo{journal}{Materials} \bibinfo{volume}{16}
  (\bibinfo{year}{2023}) \bibinfo{pages}{2067}.
  \DOIprefix\doi{10.3390/ma16052067}.
\bibitem[{Rudshaug et~al.(2023)Rudshaug, Aasen, Hopperstad, and
  Børvik}]{rudshaug_physically_2023}
\bibinfo{author}{J.~Rudshaug}, \bibinfo{author}{K.~O. Aasen},
  \bibinfo{author}{O.~S. Hopperstad}, \bibinfo{author}{T.~Børvik},
\newblock \bibinfo{title}{A physically based strength prediction model for
  glass},
\newblock \bibinfo{journal}{International Journal of Solids and Structures}
  (\bibinfo{year}{2023}) \bibinfo{pages}{112548}.
  \DOIprefix\doi{10.1016/j.ijsolstr.2023.112548}.

\end{thebibliography}
\end{document}